# Variant interpretation using population databases: lessons from gnomAD

Sanna Gudmundsson[1,2,3], Moriel Singer-Berk[1,3], Nicholas A. Watts[1,3], William Phu[1,2,3], Julia K. Goodrich[1,3], Matthew Solomonson[1,3], Genome Aggregation Database Consortium, Heidi L. Rehm[1,3,4], Daniel G. MacArthur[1,5,6], Anne O'Donnell-Luria[1,2,3]

1. Program in Medical and Population Genetics, Broad Institute of MIT and Harvard, Cambridge, MA, USA
2. Division of Genetics and Genomics, Boston Children's Hospital, Harvard Medical School, Boston, MA, USA
3. Analytic and Translational Genetics Unit, Massachusetts General Hospital, Boston, MA, USA
4. Center for Genomic Medicine, Massachusetts General Hospital, Boston, MA, USA
5. Centre for Population Genomics, Garvan Institute of Medical Research and UNSW Sydney, Sydney, New South Wales, Australia
6. Centre for Population Genomics, Murdoch Children's Research Institute, Melbourne, Australia

Correspondence: Anne O'Donnell-Luria, Broad Institute of MIT and Harvard, 415 Main St, Cambridge, 02142, MA, USA. E-mail: odonnell@broadinstitute.org

## GRANT NUMBERS

This work was supported by the National Human Genome Research Institute of the National Institutes of Health under award numbers UM1HG008900, U01HG011755, and U24HG011450 and the National Institute of Diabetes and Digestive and Kidney Diseases U54DK105566. The content is solely the responsibility of the authors and does not necessarily represent the official views of the National Institutes of Health.

## ABSTRACT

Reference population databases are an essential tool in variant and gene interpretation. Their use guides the identification of pathogenic variants amidst the sea of benign variation present in every human genome, and supports the discovery of new disease-gene relationships. The Genome Aggregation Database (gnomAD) is currently the largest and most widely used publicly available collection of population variation from harmonized sequencing data. The data is available through the online gnomAD browser (https://gnomad.broadinstitute.org/) that enables rapid and intuitive variant analysis. This review provides guidance on the content of the gnomAD browser, and its usage for variant and gene interpretation. We introduce key features including allele frequency, per-base expression levels, constraint scores, and variant co-occurrence, alongside guidance on how to use these in analysis, with a focus on the interpretation of candidate variants and novel genes in rare disease.

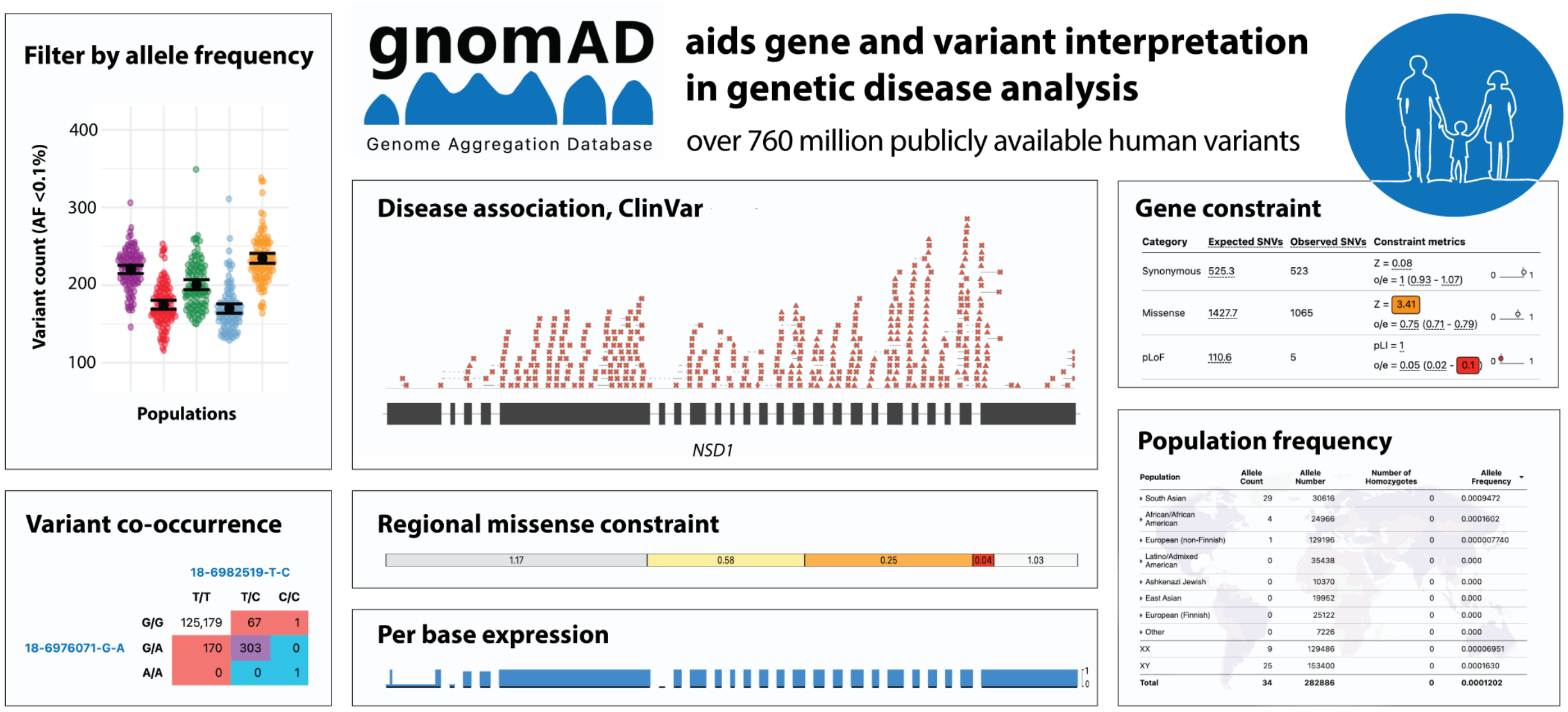

# 1. INTRODUCTION

Reference population databases are a powerful tool for understanding the biological function of genetic variation. Population frequency data allow the rare variants that are more likely to be the cause of Mendelian disorders to be distinguished from the millions of common and largely benign variants present in every human genome.

In the era prior to the availability of large sequenced cohorts, the frequency of candidate pathogenic variants was typically defined through the painstaking genotyping of small in-house cohorts of healthy individuals. However, over the last decade, a series of databases have provided increasingly more accurate and comprehensive genome-wide estimates of variant frequency through the generation and aggregation of large collections of human sequencing data. The 1000 Genomes Project was a pioneer in creating a publicly available reference database of variation from sequence data (1000 Genomes Project Consortium et al., 2010), followed by the Exome Sequencing Project, where 6,500 European and African American individuals were sequenced and aggregate data was shared on the Exome Variant Server (Fu et al., 2013). The need for a larger and more diverse reference population database was well recognized (MacArthur et al., 2014), and the first large-scale aggregation of existing sequence data from 60,000 individuals, the Exome Aggregation Consortium (ExAC) dataset, was released in 2014 (Lek et al., 2016). With the addition of genome data to ExAC, it was renamed as the Genome Aggregation Database (gnomAD), that today has variant data from more than 195,000 individuals and is currently the most widely accessed reference population dataset with over 150,000 weekly page views (Figure 1) (Karczewski et al., 2020). Other large databases include NHLBI's Trans-Omics for Precision Medicine (TOPMed)-BRAVO and the Geisinger Healthcare System DiscovEHR dataset (Dewey et al., 2016; Taliun et al., 2021).

Given known mutation rates, it is almost certain that every possible single base change compatible with life exists in a living human. Synonymous variation is under less selective pressure than missense or loss of function (LoF) variation and can be used to estimate how close gnomAD is to sampling the full spectrum of natural human variation. As of now, gnomAD is approaching saturation for the highly mutable CpG dinucleotides

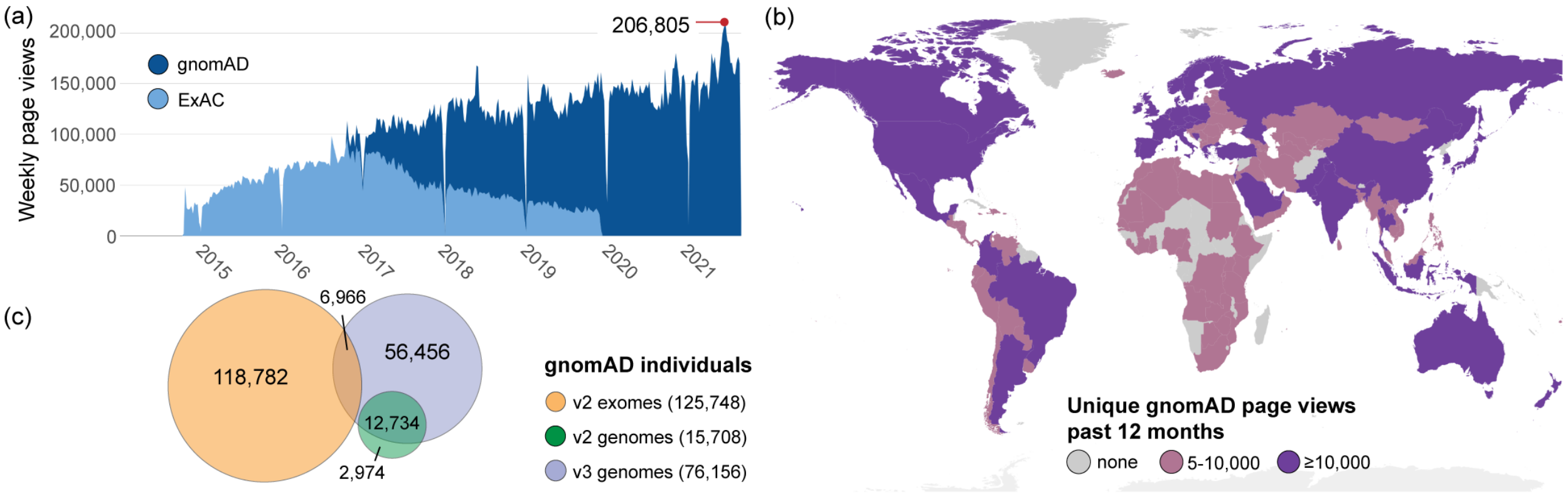

**Figure 1:** The gnomAD database aids variant interpretation world-wide. (a) Weekly page views of gnomAD (dark blue) and ExAC (light blue) from release in October 2014 to mid-2021. (b) Number of unique gnomAD page views in each country the past 12 months (since 2020-06-14) colored by none (grey), 5-10,000 (pink), and more than 10,000 (purple). (c) Schematic of the distribution and overlap of more than 195,000 unique individuals in gnomAD v2 exomes (orange), v2 genomes (green) and v3 genomes (violet).

---

(Duncan & Miller, 1980; Lander et al., 2001), with 85% of all possible synonymous CpG-to TpG transitions observed (Karczewski et al., 2020). However, across non-CpG trinucleotide contexts, less than 12% of possible synonymous variants have been observed in gnomAD, indicating that a much larger number of individuals will need to be sequenced before we begin to discover the full spectrum of tolerated variation. The fraction of observed variants is even lower for variants under purifying selection, with less than 4% of nonsense variants currently observed in gnomAD.

With the existing sample size of gnomAD, an individual will on average carry about 200 very rare coding variants (gnomAD allele frequency < 0.1%). This number varies by ancestry, partly depending on the populations represented in the database, but is also influenced by the heterozygosity rate. At the current size, each individual has tens of variants that are absent from gnomAD. Within one individual's exome, there is a mean of 27 ±13 novel coding variants that are absent in all other gnomAD individuals (variants unique to that individual), with more novel variants in East Asians (35 ± 11), and South Asians (38 ± 14) and fewer in African/African Americans (21 ± 7), Latino/Admixed Americans (19 ± 11) and Europeans (23 ± 11) (Figure 2B and Supp.

Table S1), correlating with sample sizes and representation in the dataset for each continental population. For population-groups not well represented in gnomAD, we would expect these numbers to be even higher. Often there is limited evidence to rule out pathogenicity for the variants not observed in a population database, resulting in an increased number of variants of uncertain significance in clinical genetic testing and highlighting the need for continued aggregation of sequencing data to improve the accuracy of genetic test interpretation (Naslavsky et al., 2021). The gain from increased sample size and improved representation is demonstrated by the decrease in number of unique variants per individual when utilizing the entire gnomAD dataset versus v2

---

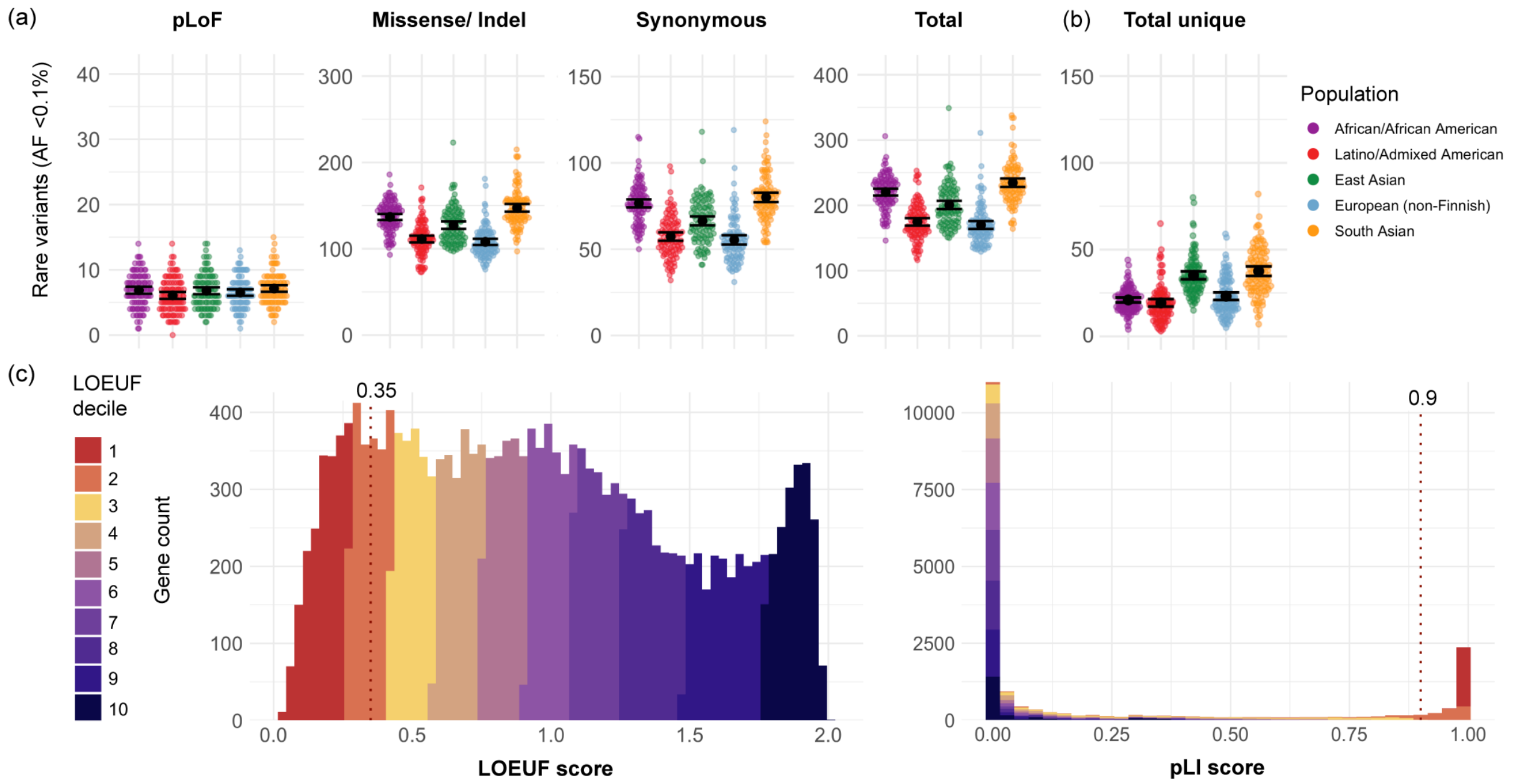

**Figure 2:** (a) Mean count of coding very rare variants (allele frequency < 0.1%), and (b) mean count of unique coding variants (across v2 and v3) grouped by population; black bar represents the 95% confidence interval. (c) Comparison of genome-wide distribution of loss of function (LoF) constraint scores in 19,197 genes, colored by LOEUF decile; a continuous distribution for LOEUF score and a dichotomous-like distribution for pLI scores. Dotted line marks suggested thresholds for LoF constrained genes at pLI ≥ 0.9 and LOEUF < 0.35 in gene interpretation.

exomes only (Supp. Figure S1 and Supp. Table S2).

This review provides guidance for using the gnomAD browser and key features like allele frequency, per-base expression levels, constraint scores and variant co-occurrence, for variant and gene interpretation in clinical and research analysis.

## 2. DATA COMPOSITION

### 2.1 Individuals represented in gnomAD

The gnomAD browser displays summary statistics and aggregate variant data from de-identified exome and genome data, with approaches consistent with the guidance in the NIH data sharing policy (NOT-OD-03-032) to reduce the risk of subject identification. The gnomAD database aggregates data from over 195,000 individuals through a world-wide collaborative effort on data sharing. More than 140 principal investigators have contributed genome data from over 60 studies, including data from several other population datasets (https://gnomad.broadinstitute.org/about). Most of the sequence data in gnomAD is generated for case-control studies of common adult-onset disease, such as type 2 diabetes, psychiatric disorders, and cardiovascular disease. No sequencing has been done for the purpose of depositing data in gnomAD; some data has been reprocessed for inclusion, particularly sequence data from populations that are under-represented in gnomAD. Data contributions were made with an assumption that no phenotype or individual-level data would be shared with users. However, access to many datasets included in gnomAD are available through the Database of Genotypes and Phenotypes (dbGaP) and the Analysis, Visualization, and Informatics Lab-Space (AnVIL) as well as other repositories (Schatz et al., 2021; Tryka et al., 2014).

Aggregation of data from disparate sources and platforms has been made possible by uniform joint variant calling using a standardized BWA-Picard-GATK pipeline (Van der Auwera et al., 2013) and Hail for data processing, analysis, and the addition of a gVCF combiner used in the v3 dataset. The aggregated dataset has been subjected to thorough sample and variant quality control (QC), with samples removed if they have low coverage, too many or too few variants for the population, or sex aneuploidy. To avoid inflation of allele frequencies for rare variants, first and second degree relatives have also been removed. In addition, to create a dataset as close as possible to a general population reference, individuals known to be affected with severe pediatric disease, as well as their first degree relatives, are also excluded. An allele-specific random forest approach (Karczewski et al., 2020) or the allele-specific version of GATK Variant Quality Score Recalibration (VQSR) have been applied to distinguish true genetic variants from artifacts. Additionally, variants were removed if no sample harboring the variant had a high quality genotype (depth ≥ 10, genotype quality ≥ 20, minor allele fraction ≥ 0.2 for non-reference heterozygous variants). With this design, gnomAD is particularly suitable for aiding in the interpretation of variants in rare disease genetic analysis.

### 2.2 gnomAD version 2 and version 3

As new cohorts are added, new versions of gnomAD are released. To date, the database consists of two versions, v2.1.1 and v3.1.2 (referred to as v2 and v3) released after the original ExAC database that is largely represented within gnomAD v2 as well as a separate dataset on the browser (Figure 3:1). With 125,748 exomes and 15,708 genomes aligned to GRCh37, gnomAD v2 is preferable over v3 for interpreting coding variants. The current v3 release has 76,156 genomes aligned to GRCh38, providing more data for noncoding regions or coding regions not covered well in exomes, such as regions with high GC content or regions not targeted with exome capture.

While gnomAD does not contain duplicated individuals, or first or second degree relatives within a version release, there is significant overlap between v2 and v3, which is important to note if using both versions for variant interpretation. The majority of v2 genomes (81%) are also in v3, additionally 6% of v2 exomes are also represented as genomes in v3. In total, approximately 14% of the individuals with exome or genome data in v2 have genome data in v3; 26% of individuals in v3 are also present in v2 (Figure 1C). The overlap of individuals can be resolved on the browser by looking at the v3 non-v2-dataset (further explored in section 3.2.1), important for rare disease analysis that benefits from investigating the entire gnomAD dataset (Supp. Figure S1 and Supp. Table S2). The planned v4 release will include the exomes and genomes from v2 and

v3, along with additional data for an expected database of over 500,000 samples aligned on GRCh38, which will be the recommended reference dataset for all analyses.

### 2.3 Structural and mitochondrial variants

This review focuses on the interpretation of single nucleotide variants (SNVs) and indels from the nuclear genome. The gnomAD browser also provides allele frequencies for structural variants (SVs) and mitochondrial variants. As part of gnomAD v2, there are annotations for ~445,000 SVs from 10,738 genomes (Collins et al., 2020) that can be explored using the search bar menu on the landing page or the gene page (Figure 3:1 and 3:2). Mitochondrial variants are available from 56,434 gnomAD v3 genomes that can be found by searching “MT-” followed by gene name or “M-” followed by a mitochondrial chromosome position. The release includes mitochondrial specific data such as homoplasmic and heteroplasmic calls as well as both population and haplogroup-specific allele frequencies (Laricchia et al., 2021).

## 3. NAVIGATING THE gnomAD BROWSER

### 3.1 The gene page

Entering a gene in the search bar (Figure 3:1), or selecting the gene name on the variant page, navigates to the *gene page* (Figure 3). The mean depth of coverage of a gene guides the first assessment of the gene's representation in the database and reveals differences between exome (Figure 3:3, blue) and genome (Figure 3:4, green) sequencing coverage. The gene displayed is a union of all exons from all transcripts. Figure 3 shows the *NSD1* gene page; haploinsufficiency of *NSD1* (loss of one copy) results in Sotos syndrome, characterized by overgrowth and intellectual disability, OMIM #117550 (Hamosh, Scott, Amberger, Bocchini, & McKusick, 2005). The direction of the arrow corresponds to the strand direction of the DNA, in this case indicating that *NSD1* is present on the forward strand (Figure 3:5). Selecting “Show transcripts” (Figure 3:6), expands the view of transcript models and any transcript can be selected for view on a separate *transcript page*. The asterix (*) marks the canonical transcript which is the Ensembl canonical transcript for v2 on GRCh37 and Matched Annotation from NCBI

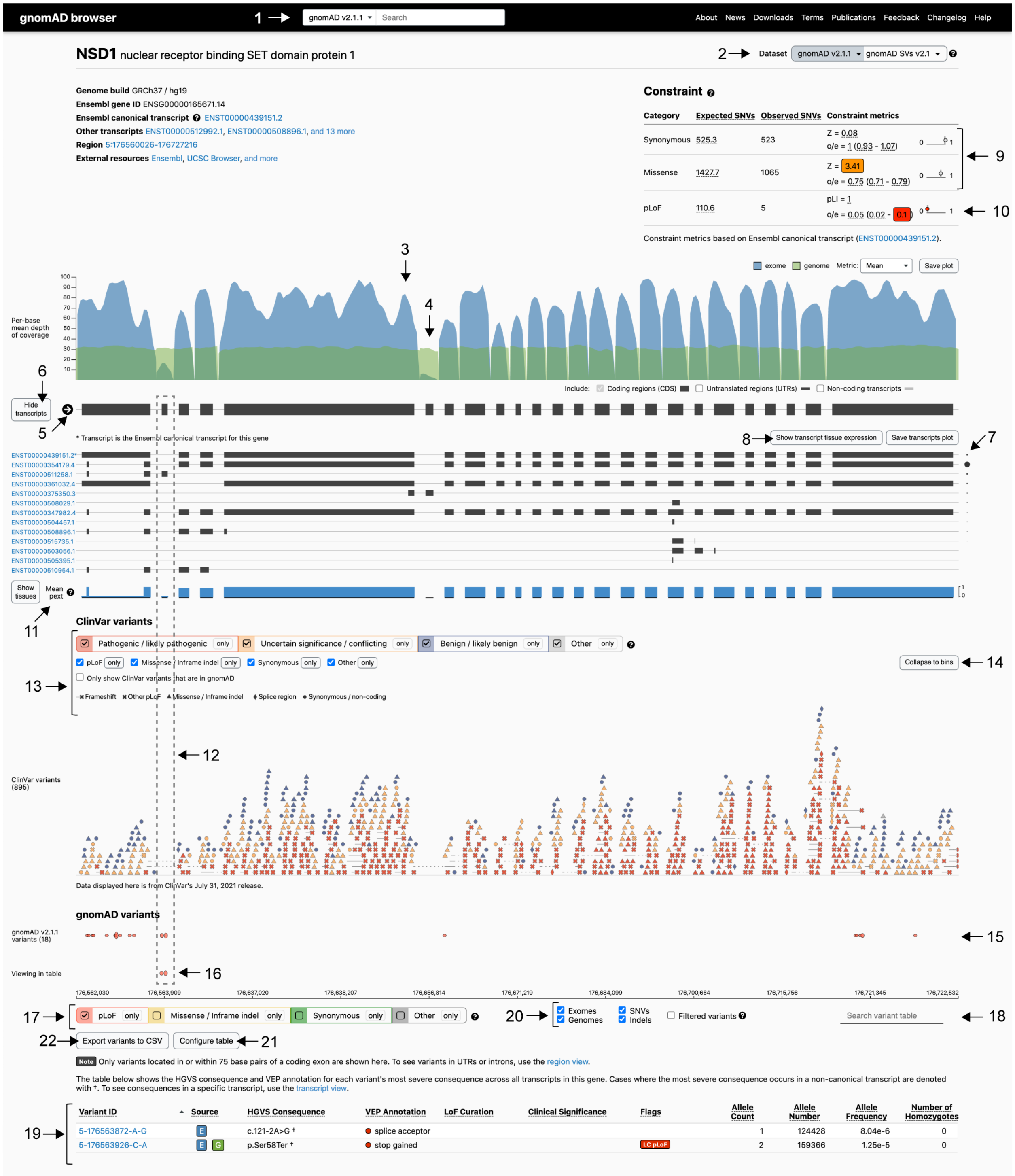

**Figure 3:** The gnomAD gene page, displaying *NSD1* as an example. Includes gene-level information of metrics and variant distribution, and allows customized filtering. Some highlighted features are: (1-2) navigating datasets; (3) exome and (4) genome gene coverage; (5) direction of the gene (*NSD1* on forward strand); (6-8) transcript and expression information; (9-10) constraint table; (11) proportion expressed across transcripts (pext) score and (12) example of a region with low pext; (13) filtering options

for ClinVar variants and (14) expansion of ClinVar variant view; (15-16) gnomAD variant tracks and (17) filter gnomAD variants by consequence; (18) variant search bar; (19) variant table; (20) filter by sequencing method, variant type, and option to include low-quality filtered variants; (21) customize variant table; (22) download variant table.

---

and EMBL-EBI (MANE) Select transcript if available for v3 and later releases on GRCh38. We display the transcripts and tissue expression profiles using data provided by the Genotype-Tissue Expression (GTEx) project, a resource for gene expression and regulation with RNA-sequencing of samples from 54 non-diseased tissue sites across nearly 1,000 individuals (GTEx Consortium, 2015). The transcript with the highest mean expression can be inferred from the size of the dot placed next to each transcript (Figure 3:7). In this example, the second listed *NSD1* ENST00000354179.4 transcript has the highest mean expression across all tissues, with the highest tissue specific expression in “Brain - Cerebellar Hemisphere”. More detail is provided in the heatmap of transcript tissue expression (Supp. Figure S2), found by selecting “Show transcript tissue expression” (Figure 3:8).

*3.1.1 Gene constraint metrics*

Constraint metrics are key features of the gnomAD database and have been widely used to aid gene and variant interpretation in rare disease (Bamshad, Nickerson, & Chong, 2019; Oved et al., 2020). The constraint metrics are based on a gene’s observed vs. expected number of very rare SNVs (allele frequency < 0.1%), corrected for sequence context and coverage. Constrained genes have fewer variants than expected and are under a higher degree of selection than less constrained genes. The metrics are divided by mutation class: synonymous, missense, and predicted loss of function (pLoF). The constraint scores for the canonical transcript are shown in the constraint table on the gene page; additionally, an updated transcript-specific constraint table is available for each transcript on the *transcript page*.

The degree of synonymous and missense constraint is measured by an observed to expected ratio (o/e) and the Z-score (Figure 3:9). A positive Z-score indicates fewer variants observed than expected, hence increased constraint (intolerance to variation), and a negative Z-score indicates that a gene has more variants observed than expected. Synonymous variants are classically not under evolutionary selective pressure indicated by a Z-score close to zero or an o/e ratio close to one. However, genes with highly similar paralogs, pseudogenes, or segmental duplications may have mapping challenges with short read data that can inflate observed variant counts. Hence a Z-score for synonymous variation deviating significantly from zero may indicate that the constraint measurement is unreliable for that gene and interpretation of all constraint data for the gene should be done with care. As an example the *HIST1H4E* gene has a synonymous Z-score of -9.89. In addition to missense constraint, some genes will have regions of missense constraint, often overlapping with functionally important domains (Samocha et al., 2017). The regional missense constraint track can be viewed by selecting the ExAC subset on the gene page. Of note, the track is only shown for genes where regional missense constraint is observed (Figure 3:2, Supp. Figure S2).

LoF constraint is measured by two different scores: probability of being LoF intolerant (pLI) and LoF observed/expected upper bound fraction (LOEUF) (Figure 3:10). The pLI score was developed for the ExAC release (Lek et al., 2016) on the premise that genes can be divided into genes where biallelic LoF is tolerated by natural selection (LoF tolerant genes, pLI close to 0) and genes where LoF is not tolerated (haploinsufficient genes, pLI close to 1) (Figure 2C, right). However, the increased sample size of gnomAD enhanced the power to detect LoF constraint with a continuous metric; LOEUF is a conservative estimate (upper bound) of the observed/expected ratio of the LoF effect predictor (LOFTEE) high confidence SNV-LoF variants (Figure 2C, left). The use of constraint score in analysis is further explored under section 4.2.

### *3.1.2 Expression score (pext)*

Using per-base expression annotations aids in determining if a variant occurs in a biologically relevant exon and helps deprioritize variants unlikely to impact gene function (Abou Tayoun et al., 2018). The gnomAD browser displays the mean proportion expressed across transcripts (pext) score, based on GTEx v7 data from 800 individuals (GTEx Consortium, 2015). The pext score is unique in providing a normalized expression value for each position in a gene. By doing so, pext allows quick visualization of the mean expression of exons across a gene either as an aggregate score including all tissues (Figure 3:11, blue track), or separated across 38 different tissues by selecting “show tissues” (Cummings et al., 2020). The *NSD1* gene page demonstrates the applicability of pext by occurrence of pLoF variants in gnomAD individuals in low pext regions and, to a larger extent, a lack of pathogenic ClinVar variants in the same regions (Figure 3:12). Of note, pext is based on adult postmortem tissue and may not accurately represent genes with differential expression during development (GTEx Consortium, 2015). The use of pext in variant interpretation is explored under section 4.3.

### *3.1.3 Variant tracks and the variant table*

The gene-based overview of ClinVar variant distribution can be viewed on the gene page (pathogenic/likely pathogenic in red, variants of uncertain significance and conflicting interpretation in orange, and likely benign/benign in blue). This track shows all coding ClinVar variants (and 75 base pairs surrounding the exons) irrespective of whether they are found in gnomAD, with an option to view only ClinVar variants present in gnomAD (Figure 3:13). By expanding the ClinVar track (Figure 3:14), the clinical significance of each variant will be indicated by shape. This allows for visualization of patterns that can inform variant classification such as identifying genes where there is a predominant type of pathogenic variation (LoF or missense variation) or regional clustering of missense variation (hot spot). *NSD1* (Figure 3) displays clear enrichment of pathogenic (red) pLoF (X) variants, consistent with haploinsufficiency of *NSD1* resulting in Sotos syndrome. There are also pathogenic missense variants (red triangles) that cluster within a portion of *NSD1* with evidence of regional missense constraint, harboring the SET and PWWP domains (Supp. Figure S3).

The gnomAD variant track shows the distribution of variants along the gene, with the height of the oval scaled by allele frequency (Figure 3:15 and 3:16). Underneath this track there are filtering options for visualizing one or more types of variants (Figure 3:17). The upper track displays the variants matching the selections across the exons

+/- 75 surrounding base pairs, in this example, filtered to only include pLoF variants in *NSD1* (Figure 3:15); the lower track displays the variants visible in the variant table to help with orientation within the gene (Figure 3:16). Further, a variant of interest can be highlighted using the search box (Figure 3:18), which assists inspection of pext score and ClinVar variants overlapping the gnomAD variant of interest.

The variant table lists all variants in a gene sorted by genomic position (Figure 3:19, Variant ID). The most severe consequence across transcripts is noted in the HGVS Consequence column. When the most severe consequence occurs in a transcript that is non-canonical, the HGVS nomenclature is denoted with a † symbol. The variants in the table can be filtered by variant effect, exomes or genomes, and SNV or indels, and there is also an option to include variants that did not pass gnomAD quality control; however, these variants should be interpreted with caution (Figure 3:20). Configuring and re-ordering the variant table (Figure 3:21), and sorting by column values of interest is also possible, for example including pext score for each variant site. Some pLoF variants will have a LoF curation verdict and warning flags noted in the table (Supp. Figure S4). More information about LOFTEE and the flags it generates as well as the manual LoF curation results that are present for a subset of variants and genes, can be found by hovering over these flags, or on the variant page. Furthermore, the customized variant table can be downloaded as a comma separated values (CSV) file for local analysis (Figure 3.22).

### 3.2 The variant page

Searching for a variant in the search bar (Figure 3:1) or selecting a variant in the variant table, navigates to the *variant page*. The variant page displays annotations and summary statistics for a specific variant as well as links to external resources reporting information on that variant including dbSNP (Sherry et al., 2001), UCSC (Kent et al., 2002) and ClinVar (Landrum et al., 2018) (Figure 4:1). Any concerns or questions regarding the validity of a specific variant or its annotations can be submitted via the feedback link (Figure 4:2), such as a variant that passes QC but appears to be a sequencing artifact or variants though to be associated with severe, penetrant, early-onset dominant disease.

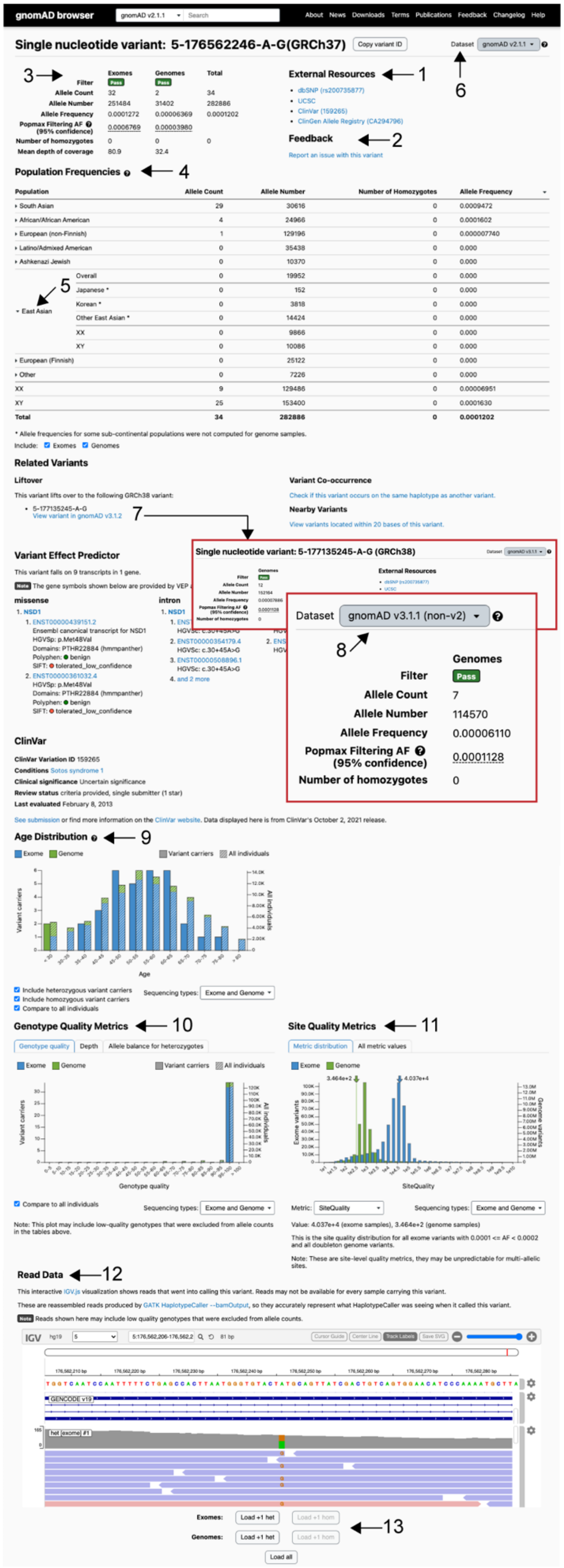

**Figure 4:** The gnomAD variant page, displaying the *NSD1* missense variant 5-176562246-A-G, p.Met48Val (NM_022455.5:c.142A>G) as an example. Includes variant level information and site specific metrics. Some highlighted features are: (1) external resources and (2) variant feedback forms; (3) allele frequency summary table with filtering allele frequency; (4) population frequency table and (5) visualization of subcontinental populations; (6) navigating datasets; (7) liftover link for gnomAD v3 and (8) visualization of v3 non-v2 dataset; (9) age data; (10) genotype/depth/allele balance for heterozygotes and (11) site quality metrics; (12) read data and (13) the option to load read data for additional individuals.

*3.2.1 Allele frequency and allele count*

Display of robust allele frequencies across the database (global) and within continental populations is a main feature of gnomAD. An overview of the allele frequencies, including the filtering allele frequency (FAF), of a variant is found at the top of every variant page together with information on variant quality control filters (Figure 4:3). The population frequency table (Figure 4:4) contains allele frequency information from five continental populations (African/African American, East Asian, European, Latino, and South Asian), two demographically distinct populations (Ashkenazi Jewish and Finnish), Middle Eastern (v3 only), and any remaining uncategorized (Other) samples. Some subcontinental populations are available (Figure 4:5) and differ between gnomAD releases. The population frequency table provides an opportunity to identify enrichment of variants within populations, in this example 29 of 34 heterozygous carriers are from the South Asian population. Allele frequency is calculated by dividing allele count by allele number, i.e. allele frequency represents the frequency of *confidently sequenced haplotypes* that carry the allele in question (because coverage and sequencing quality varies across the genome, the allele number can differ substantially between positions). Allele frequency is *not* equivalent to the percentage of individuals that carry the allele, but is a suitable value for expressing the frequency of a variant in the general population. The number of individuals carrying a variant will depend on the number of heterozygous and homozygous individuals but can be calculated from the data provided in the variant table. Of note, the allele number (number of chromosomes genotyped) is calculated on the dataset where the variant is identified, so it will be lowest for a variant that is identified in v2 genomes only and highest if present in v2 exomes and genomes. Presence in v2 genomes only is the most common reason for low allele number and it is not a cause for concern. The regional coverage should be investigated by looking at allele numbers of proximal variants in the variant table (Figure 3:19) and review of the coverage data (Figure 3:3 and 3:4).

The dataset menu (Figure 4:6) provides the opportunity to explore allele frequency in specific subsets of gnomAD (non-cancer, non-neuro, non-TOPMed, control and non-v2). The control subset consists of individuals reported as controls in common disease studies or included from a biobank, but these individuals may still have medical conditions. For example, someone who participates as a control in a type 2 diabetes study could have a past history of, or in the future develop cancer or neuropsychiatric disease and still be included in the control subset. This is one reason that these subsets should not be used as a control set in common disease studies (further discussed in section 5). While we expect individuals with severe early-onset disease to be heavily depleted from gnomAD, individuals with conditions that would not prohibit participation in common disease research are likely included at mildly depleted or similar levels to the general population, particularly for phenotypes like infertility, vision and hearing impairment, and conditions with late onset or reduced penetrance. It can be useful to investigate the allele count across the database for rare variant analysis (Supp. Figure S1 and Supp. Table S2) by using the liftover link (Figure 4:7) and the non-v2 subset in v3 to resolve any overlaps between versions (Figure 1C and Figure 4:8). In this example, in addition to the 34 carriers in v2, 12 carriers of the *NSD1* p.Met48Val (NM_022455.5:c.142A>G) variant are found in v3 of which 7 individuals are unique to v3, adding up to a total of 41 heterozygous carriers across the entire gnomAD dataset. The use of allele frequency in variant interpretation is explored under section 4.1.

*3.2.2 Age data*

Age data is available for a subset of individuals. While it is defined as the last known age, for some cohorts it is the age at enrollment (Figure 4:9). The data is displayed as a distribution including the age of carriers and the age of all individuals in the dataset (striped), colored by sequencing method (blue for exome, green for genome). For variants in genes associated with later onset conditions, age data can reveal if a carrier is of an age younger than onset of disease, which might explain the presence of a disease-variant in gnomAD. In general, age data is of limited use in variant interpretation given the lack of available phenotype data on these individuals.

*3.2.3 Quality control*

Variants reported in gnomAD have passed robust quality control including hard filters and a random forest model for assessing both the quality of the variant and the site. However, presence of artifacts is inevitable in any reference database and careful review of individual variants is important, especially in rare disease analysis. While sequence artifacts occur relatively evenly across the genome, biologically important sequences are depleted for natural variation, which results in a relative enrichment of artifacts compared to natural variation in disease-associated regions (Gudmundsson et al., 2021).

Genotype quality metrics (Figure 4:10) provide the genotype quality, read depth, and allele balance for all individuals genotyped at the site of the variant. Of note, the number of samples represented in the histograms may be discordant with the variant table as the graph includes individuals filtered by hard filters (depth < 10, genotype quality < 20, minor allele balance < 0.2 for alternate alleles of heterozygous genotypes). When examining a variant possibly associated with a severe pediatric disorder, low genotype quality may suggest the variant is an artifact or skewed allele balance might suggest the variant is an artifact or mosaic in the gnomAD individual. Quality concerns can serve as an explanation for why the variant is observed in gnomAD, and can strengthen a disease-association hypothesis, particularly if the presence of a variant is Sanger confirmed in a patient. Site quality metrics used in the random forest model, such as Fisher strand bias, are also available (Figure 4:11).

Read data is available for most variants on the browser, and examining the variant site using the Integrative Genomics Viewer (IGV) is an important part of variant evaluation (Figure 4:12 and 4:13). Read data should be reviewed for evidence of strand bias, skewed allele balance, indications of mapping issues (multiple variants in the region), drop in coverage, and nearby variants that may affect the current variants' interpretation. For detailed information on how to use IGV we refer to the IGV User Guide and previous articles (Robinson, Thorvaldsdóttir, Wenger, Zehir, & Mesirov, 2017). Not all CRAMS were still available during gnomAD production to generate read data. For variants where read data is missing from the variant page of one version of gnomAD (v2, v3, and ExAC), we suggest investigating if they are represented in another version.

## 4. VARIANT INTERPRETATION USING gnomAD

The gnomAD dataset is used in the majority of rare disease analysis pipelines in both diagnostic and research settings around the world (Figure 1B). The American College of Medical Genetics and Genomics (ACMG) and Association for Molecular Pathology (AMP) have defined standards for variant classification (Richards et al., 2015). These provide rigorous guidance for the evaluation and aggregation of variant evidence, including the use of reference databases such as gnomAD. They defined terminology for variant classification including five categories: *benign, likely benign, uncertain significance, likely pathogenic*, and *pathogenic*. Further, they defined four major areas in which variants can be awarded with evidence that determines their final classification including: *population, computational, functional*, and *segregation* data.

### 4.1 Allele frequency in variant interpretation

The vast majority of pathogenic variants are rare, hence identifying rare variants is an essential step in Mendelian analysis. It is important to remember that most rare variants are not pathogenic and rarity is consistent with, but not sufficient for, determining pathogenicity (Figure 2A and B). Variants that are absent from gnomAD, or present at a lower frequency than expected (particularly for recessive disease variants where unaffected carriers are expected), were initially considered a moderate level of evidence for pathogenicity (PM2) (Richards et al., 2015). However, rarity or absence in population databases has been recommended to be downgraded to supporting evidence by the Clin Gen Sequencing Variant Interpretation (SVI) Working Group, given that most unrepresented variation is benign and reference population databases are far from saturation for most variation types (Karczewski et al., 2020).

Population evidence for pathogenicity can also be applied for variants that are more prevalent in affected individuals compared to controls (PS4). There have been some attempts to use gnomAD as a control population in formal association studies, but in general this is not recommended. Challenges to this approach include the lack of information about case numbers for any specific disease in gnomAD, the inability to correct for confounders such as population stratification (which would require individual-level data access), as well as differences in the technical processing and QC of data between cases and controls which can lead to erroneous associations (Karczewski, Gauthier, & Daly, 2019).

To most effectively filter out common variants, we recommend using the popmax allele frequency, defined as the population maximum allele frequency in the continental populations (African/African American, East Asian, European, Latino/Admixed American, and South Asian). Generally, if a variant is common in one population, it can be assumed to be benign across all populations. Consideration should be taken if studying a condition that is much more common in one population.

Following ACMG variant classification guidelines, stand-alone benign (BA1) evidence should be applied to variants with an allele frequency ≥ 5%, unless the variant was previously noted to be pathogenic, as high allele frequency can be a result of low penetrance in monogenic disease genes (Ghosh et al., 2018). Hypomorphic variants in particular may have a higher allele frequency. An allele frequency of > 1% is considered strong evidence that the variant is benign (BS1); however, certain recessive disorders can have common pathogenic variants that rise above this threshold (e.g. Phe508del in *CFTR* associated with Cystic Fibrosis, OMIM #219700); these well-established variants can often be identified using ClinVar. There are occasions when 1% is a too conservative threshold, particularly for severe, dominant disorders. Whiffin et al. has developed a more refined frequency filtering approach (Whiffin et al., 2017). A maximum credible allele frequency for the specific condition is defined using information about the prevalence, inheritance mode, penetrance, and genetic architecture (accounting for maximum genetic or allelic contribution). As gnomAD is a sampling of the general population, a filtering allele frequency (FAF) is generated from the popmax allele frequency to adjust for sampling variance (Figure 4:3). If the FAF is higher than the maximum credible population allele frequency, then benign evidence (BS1) can be applied.

Estimating allele frequency in genes affected by clonal hematopoiesis has been a particular challenge (Carlston et al., 2017; Karczewski et al., 2020) as somatic variants that increase proliferation of the hematopoietic lineage can rise to high allele fraction (Jaiswal et al., 2014). A warning has been added to the *gene page* for genes where this phenomenon has been shown to occur (i.e. *ASXL1, DNMT3A, TET2*). Variants in these genes should be interpreted with caution in any reference population database (Jaiswal et al., 2014).

**4.2 Constraint scores in variant interpretation**

Constraint scores are useful to indicate when specific types of variation are depleted in a gene. For example, haploinsufficient genes often have a high pLI score and a low LOEUF score; therefore pLoF variants occurring within LoF-constrained genes are of high interest (Bamshad et al., 2019). Choice of score may be influenced by whether the analysis is more suited to a continuous (LOEUF) or dichotomous-like (pLI) metric (Figure 2C). When a cut off is being applied, we recommend using pLI ≥ 0.9 (3,060 genes in v2) or LOEUF < 0.35 (2,968 genes in v2). While many LoF constrained genes are associated with Mendelian disease to date, 68% (2,071 of 3,060) of LoF constrained genes (pLI ≥ 0.9) are not yet linked to a phenotype in humans.

Genes with a missense Z-score ≥ 3.1 are significantly depleted for missense variation. Unlike LoF variation, there may be only a region of a gene that is intolerant for missense variation rather than the entire gene, often overlapping with a protein domain. Reviewing the ClinVar track for pathogenic variants along with the regional missense constraint track (Supp. Figure S3) can help identify hotspots or domains without benign variation, providing moderate evidence towards variant pathogenicity (PM1) (Harrison, Biesecker, & Rehm, 2019). For missense constrained genes with many pathogenic missense variants, this can be considered supporting evidence of pathogenicity (PP2) (Harrison et al., 2019). At the other end of the missense variation spectrum lie genes without evidence of missense constraint (Z scores around zero or less) and where only truncating variants have been reported as pathogenic, which is considered as supporting benign evidence for classification (BP1) for missense variants.

In rare disease analysis, careful attention should be paid to rare variation in constrained genes, both in prioritizing variants and also in identifying novel disease-gene relationships. There are some caveats that are worth noting. Constraint is more commonly seen for dominant disease genes, particularly for phenotypes that are absent or depleted in gnomAD. It is not as informative in the interpretation of recessive disease, as carriers of recessive disease variants will be present in gnomAD. As constraint is due to negative selection, variation that results in a phenotype later in life, particularly in post-reproductive years, may not be depleted in gnomAD. For example, LoF variants in the *BRCA1* gene are strongly correlated with risk for breast and ovarian cancer, yet *BRCA1* does not show evidence of constraint for LoF variation, as the phenotype presents post-reproduction and also is of lower penetrance in males.

A full list of all constraint metrics is available on the gnomAD Downloads page (https://gnomad.broadinstitute.org/downloads#v2-constraint) including genome-wide pLI and LOEUF ranking of all genes.

**4.3 Per-base expression in variant interpretation**

Proportion expressed across transcripts (pext) scores can be used to examine the per-base expression pattern across transcripts and exons as well as in a tissue of interest (Cummings et al., 2020). Regions with low pext scores are likely of less biological importance. While the pext score is helpful in the interpretation of any coding variant, it can be particularly useful when deciding whether to apply the very strong evidence for a pLoF variant in a gene where LoF is a known mechanism of disease (PVS1). A pext score < 0.2 is evidence that a pLoF at that site may not be of biological relevance and PVS1 may not apply.

For conditions impacting specific tissues, the pext score for that tissue can be reviewed, as transcript expression between tissues can differ. Of note, pext score is based on adult tissue and the developmental expression profile may be different. Additionally, particularly long genes can be affected by the 3’ bias of polyA tail transcriptome sequencing to a degree which is not always adequately corrected for in pext scores, and genes with this pattern should be interpreted with caution (for example *DMD*). It is important to consider the relative value compared to the mean and maximum pext value of the gene. For example, if a variant falls in an exon with a pext of 0.1, but the gene average is 0.2 (such as in the setting of expression of a noncoding transcript), the data does not provide any information for interpretation, and PVS1 may still apply.

## 4.4 Variant co-occurrence

The gnomAD variant co-occurrence feature allows investigation of the statistical likelihood of two variants occurring on the same or different haplotypes in individuals in gnomAD. This can be helpful in rare disease analysis to deprioritize compound heterozygous genotypes that are seen in the general population or to predict phasing where only a proband is available for sequencing and two rare variants are observed within a gene. The co-occurrence can be assessed if both variants are in gnomAD exomes, appear in the same gene, have an allele frequency ≤5%, and are coding, in the splice region, or in the 5' or 3' untranslated regions (UTRs). When two variants co-occur in individuals in gnomAD more often than expected based on population frequencies, then the variants are predicted to be on the same haplotype (Supp. Figure S5A). If two variants co-occur in some individuals in gnomAD but only at the expected rate, then these variants are predicted to be on different haplotypes, and it is unlikely that this variant combination is deleterious for phenotypes that are not expected to be seen in gnomAD (Supp. Figure S5B). If two variants are found in gnomAD but do not co-occur in any individuals in gnomAD, then they are likely on different haplotypes but no information would be available about the potential impact of a compound heterozygous genotype. These estimates will be more accurate when considered within a specific population. A detailed description of the varant co-occurrence feature is found in the News section (https://gnomad.broadinstitute.org/news).

## 4.5 gnomAD flags and warnings

The gnomAD browser provides flags and warnings displayed in the variant table and on the variant page to highlight variant details important for interpretation. Variants with flags or warnings should not be automatically discounted but we advise careful consideration of whether these may impact the analytical validity or the effect of the variant.

### *4.5.1 Multi-nucleotide variants*

When two variants in the same codon are present on the same strand, the variant consequence may be misrepresented as the type if each variant were present independently rather than interpreting the variant combination. The assumption of independence is present in almost all current variant annotation pipelines. Multi-nucleotide variants (MNVs) that occur within a codon are annotated in gnomAD v2 data to aid interpretation of their combined effect (Wang et al., 2020). For example, two variants independently annotated as nonsense and missense could result in a missense variant when interpreted in combination (Supp. Figure S6). Specific information about the MNVs identified in gnomAD can be found on the MNV page, including the number of individuals who have the MNV versus either of the SNVs.

Frame restoring indels (for example a 4 base pair deletion and nearby 5 base pair deletion on the same haplotype) are *not* annotated in the gnomAD browser. However, these are an important source of rescues and can be identified by inspecting the read data (Supp. Figure S7).

*4.5.2 Loss of function transcript effect estimator (LOFTEE) and manual LoF curation*

The LOFTEE package was developed to complement pLoF annotations by variant effect predictor (VEP) and filter variants that are unlikely to result in LoF. Variants are determined as low-confidence if predicted to not result in LoF due to criteria such as terminating at the 3' end of a gene or affecting splicing of the UTR. The remaining variants are categorized as high-confidence (Karczewski et al., 2020). Around 14% of high-confidence pLoF variants in gnomAD have additional flags (i.e. single exon genes, poor conservation by Phylogenetic Codon Substitution Frequencies (PhyloCSF) (Lin, Jungreis, & Kellis, 2011), NAGNAG splice acceptor sites, and non-canonical splice sites) that can be found on the variant page together with the VEP information. The additional flags differ from those variants that are filtered as low-confidence pLoF as they generally relate to the properties of individual transcripts or exons and can be overruled by gene-specific knowledge.

For a subset of pLoF variants, manual curation of the effect of pLoF variants has been performed. The results are displayed below the VEP annotations on the variant page and in the variant table on the gene page (Supp. Figure S4). The flags and manual curation verdicts will often impact whether PVS1 can be applied; however, careful review is essential in determining if these flags change the interpretation of the variant in a specific context. Of note, pLoF variants suggested to not result in LoF or pLoF variants with flags could still be pathogenic, as LoF might not be the only mechanism of disease.

*4.5.3 Low complexity regions*

Low complexity sequences are enriched for artifacts. The low complexity region (LCR) flag highlights variants identified in these regions to allow a more careful review (Morgulis, Gertz, Schäffer, & Agarwala, 2006). The allele frequency of variants in LCRs might be skewed because of enrichment for artifacts, and population frequency evidence (PM2/BA1/BS1/BS2) should be applied cautiously. An additional type of LCR that is often not flagged in gnomAD are homopolymer runs, which also show an enrichment of sequence artifacts. Skewed allele balance can provide evidence that the variant is an artifact in the population data. However, pathogenic variants also commonly occur here. If the variant is Sanger confirmed in the patient, it should not be discounted as causal despite being present at an apparent appreciable frequency in population data.

*4.5.4 Variant page warnings*

There are two main warnings that may be seen on the variant page underneath the header. Firstly, variants that are covered in less than 50% of individuals are highlighted since there is risk that variant allele frequencies may be inaccurate (Supp. Figure S8). These variants often fall in regions that are difficult to sequence with current exome and genome methods. Coverage can also vary because of differences in exon capture between different sequencing platforms. If a variant is better covered in the genome data, using v3 allele frequencies might be more appropriate.

Secondly, heterozygous variants with a high proportion of alternate reads (allele balance for the variant is ≥ 90%) are highlighted (Supp. Figure S9). Heterozygous variants with inflated alt-reads are likely homozygous variants called heterozygotes due to contamination that affects the variant calling likelihood models. This depletion of homozygous calls can incorrectly deflate the allele frequencies (Karczewski et al., 2019).

## 5. LIMITATIONS OF REFERENCE POPULATION DATABASES

The gnomAD database is a useful, publicly available collection of human sequence data, but there are a number of caveats that are important to note when drawing inferences about variant pathogenicity from this resource (or, indeed, most other existing variant databases).

It is important to note that some individuals with Mendelian disease may still be included in the datasets. We suggest caution about excluding variants as disease candidates when seen in one or a few individuals. Also, as demonstrated in Figure 2A, 2B, and prior work (Karczewski et al., 2020), we are still far from representing all possible variation, and a variant's absence from gnomAD is consistent but far from sufficient evidence for its involvement in disease (Figure 2B).

Phenotype and other individual-level data is not available for individuals included in the aggregate gnomAD data. This is better accessed through a biobank or other study, such as the UK Biobank, BioMe, FinnGen, or All of Us. Because individuals known to have Mendelian phenotypes have been removed, gnomAD would not be useful to try to match for patient phenotype. Other resources for this type of matching are available, including Genotype 2 Mendelian Phenotype (Geno2MP) and VariantMatcher, which are aggregated databases of rare disease sequence data with associated HPO terms and the ability to contact the researcher.

The over-representation of European participants in genetics studies is reflected in the database. There is poor representation of many communities, including African, Middle Eastern, and Oceanian populations, leading to patients from these communities having more rare variants of uncertain significance. Improving diverse representation of populations is of high priority, with resources dedicated to reprocessing available datasets for inclusion in gnomAD.

Despite extensive quality control, gnomAD (like any large genomics resource) contains sequencing and annotation artifacts, and we have suggested approaches to evaluate variant quality.

## 6. RESOURCES

Additional features are frequently released on the gnomAD browser. Extensive feature releases are described in the News section (https://gnomad.broadinstitute.org/news) and updates to the browser are detailed in the Changelog (https://gnomad.broadinstitute.org/news/changelog). Additionally, updates are announced by @gnomAD_project on Twitter.

On the browser, additional information is available via the "?" buttons located throughout the gnomAD pages and on the Help page (https://gnomad.broadinstitute.org/help), or the team can be contacted directly at gnomAD@broadinstitute.org. Further, the UCSC genome browser has a gnomAD track that allows interactive view of gnomAD metrics, including features such as allele frequency, constraint scores, and pext.

For deeper understanding about the gnomAD dataset beyond what is covered in this review, primary gnomAD publications are open access and listed on the Publications page on the browser. A complement to this review is a video tutorial on using the gnomAD browser presented at the H3Africa ClinGen Rare Disease Workshop (https://youtu.be/XdjjHdiVlrE, February 2021) or the Broad Institute Primers on Medical and Population Genetics (available on the Broad Institute's YouTube channel).

## 7. CONCLUDING REMARKS

The gnomAD resource illustrates both the power and the challenges of interpreting human biology using large-scale genomic datasets. The sheer size of gnomAD makes it possible to obtain accurate estimates of allele frequency extending down to incredibly rare variation, and to explore the patterns of variation across genes and regions of genes. This power has proven invaluable for variant interpretation in patients with rare genetic disorders, and has also empowered a wide range of scientific applications including comparison of the mutational intolerance of mouse and human genes (Dickinson et al., 2016), estimation of selection coefficients (Cassa et al., 2017), assessment of the relative evidence for reported disease genes (Walsh et al. 2017), and determination of the penetrance of dominant disease variants (Minikel et al. 2016). The gnomAD database has also aided in the discovery of genes associated with many diseases, including neurodevelopmental and congenital heart disorders (Kosmicki et al. 2017; Deciphering Developmental Disorders S...; Jin et al. 2017; Kaplanis et al. 2020).

Future releases of gnomAD will further increase the size and scope of the resource, leading to improved power for all downstream applications. New releases of structural variation calls will increase the resolution of analysis for large deletions, duplications, inversions, and complex rearrangements, while the next exome release (expected to exceed 500,000 individuals, mapped to GRCh38) will dramatically enhance power for assessing coding allele frequency as well as constraint against gene disruption and regional missense variation. Over time we expect that improvements in variant-calling methods for currently under-represented classes of variation, such as repeat expansions and complex structural rearrangements, will provide increasingly accurate frequency estimates for these variants and support the discovery of additional pathogenic alleles.

Finally, a major ongoing focus on increasing the representation of diverse ancestries, both from the gnomAD aggregation effort, and from the broader human genomics community, will be needed to improve the applicability of this database to currently under-represented populations. This effort will require greater efforts to ensure these communities are included in global genomics projects, as well as ensuring that the resulting data are shared with aggregation efforts in a manner that balances accessibility with respect for the wishes of communities and individuals, especially for Indigenous peoples (Hudson et al., 2020). Increased representation of all communities will decrease the number of variants of uncertain significance in patients from currently under-represented ancestries, while also improving the power of this resource for all communities.

## ACKNOWLEDGMENTS

We thank the individuals whose data is in gnomAD for their contributions to research. We thank Ellie Seaby for generous data sharing and inspiration for Figure 2C (Seaby,

Rehm, & O’Donnell-Luria, 2021) and Kaitlin Samocha for guidance on constraint interpretation. Development of the Genome Aggregation Database was supported by NIDDK U54 DK105566 and the NHGRI of the National Institutes of Health under award number U24HG011450. Development of the gnomAD browser for rare disease interpretation is supported in part by National Human Genome Research Institute UM1HG008900 to H.L.R., D.G.M., and A.O-D.L.; U01HG011755 to A.O-D.L. and H.L.R.; U24HG011450 to H.L.R.; and the National Institute of Diabetes and Digestive and Kidney Diseases U54DK105566 to D.G.M. The content is solely the responsibility of the authors and does not necessarily represent the official views of the National Institutes of Health. S.G. was supported by The Wallenberg Foundation scholarship program for postdoctoral studies at MIT and Broad Institute.

## DATA AVAILABILITY STATEMENT

The gnomAD data is displayed on the browser https://gnomad.broadinstitute.org/, available for download on https://gnomad.broadinstitute.org/downloads through Google Cloud Public Datasets, the Registry of Open Data on AWS, Azure Open Datasets, and the UCSC genome browser. New features and information on the gnomAD browser are shared in https://gnomad.broadinstitute.org/news/changelog, https://gnomad.broadinstitute.org/help, and https://gnomad.broadinstitute.org/news/. The gnomAD data are in part based on data that are available in TOPMed or dbGaP including: (1) generated by The Cancer Genome Atlas (TCGA) managed by the NCI and NHGRI (accession: phs000178.v10.p8); information about TCGA can be found at http://cancergenome.nih.gov; (2) generated by the Genotype-Tissue Expression Project (GTEx) managed by the NIH Common Fund and NHGRI (accession: phs000424.v7.p2); (3) generated by the Alzheimer's Disease Sequencing Project (ADSP), managed by the NIA and NHGRI (accession: phs000572.v7.p4). For a full list of projects included in gnomAD, please see https://gnomad.broadinstitute.org/about.

## CONFLICT OF INTEREST

DGM is a founder with equity of Goldfinch Bio, and serves as a paid advisor to GSK, Variant Bio, Insitro, and Foresite Labs. AO-DL is on the Scientific Advisory Board for Congenica.

## WEB RESOURCES

ClinGen SVI https://clinicalgenome.org/working-groups/sequence-variant-interpretation/
UCSC genome browser gnomAD tracks GRCh37 http://genome.ucsc.edu/cgi-bin/hgTrackUi?hgsid=1198963947_L5aQu0WaVilB3n3k3QPWX9OuYAOV&db=hg19&c=chr6&g=gnomadSuper
UCSC genome browser gnomAD tracks GRCh38 http://genome.ucsc.edu/cgi-bin/hgTrackUi?hgsid=1198963947_L5aQu0WaVilB3n3k3QPWX9OuYAOV&db=hg38&c=chr6&g=gnomadVariants

gnomAD links in the manuscript

https://gnomad.broadinstitute.org/about

https://gnomad.broadinstitute.org/

https://gnomad.broadinstitute.org/downloads#v2-constraint

https://gnomad.broadinstitute.org/news

https://gnomad.broadinstitute.org/news/changelog

https://gnomad.broadinstitute.org/help

https://gnomad.broadinstitute.org/publications

https://gnomad.broadinstitute.org/downloads

NIH Statement on Sharing Research Data https://grants.nih.gov/grants/guide/notice-files/not-od-03-032.html

Hail https://hail.is/

ClinVar https://ncbi.nlm.nih.gov/clinvar/

IGV user guide https://software.broadinstitute.org/software/igv/UserGuide

gnomAD educational video https://youtu.be/XdjjHdiVlrE

# SUPPLEMENTARY DATA: Variant interpretation using population databases: lessons from gnomAD

Sanna Gudmundsson[1,2,3], Moriel Singer-Berk[1,3], Nicholas A. Watts[1,3], William Phu[1,2,3], Julia K. Goodrich[1,3], Matthew Solomonson[1,3], Genome Aggregation Database Consortium, Heidi L. Rehm[1,3,4], Daniel G. MacArthur[1,5,6], Anne O'Donnell-Luria[1,2,3]

1. Program in Medical and Population Genetics, Broad Institute of MIT and Harvard, Cambridge, MA, USA
2. Division of Genetics and Genomics, Boston Children's Hospital, Harvard Medical School, Boston, MA, USA
3. Analytic and Translational Genetics Unit, Massachusetts General Hospital, Boston, MA, USA
4. Center for Genomic Medicine, Massachusetts General Hospital, Boston, MA, USA
5. Centre for Population Genomics, Garvan Institute of Medical Research and UNSW Sydney, Sydney, New South Wales, Australia
6. Centre for Population Genomics, Murdoch Children's Research Institute, Melbourne, Australia

## CONTENT

# SUPPLEMENTARY METHODS

## Analysis of very rare coding variants in gnomAD populations

A sample of 100 individuals was randomly selected from each of the following populations in the gnomAD v2.1.1 exome dataset: African/African American, Latino/Admixed American, East Asian, Non-Finnish European, and South Asian. Variants were then filtered to only very rare variants using a popmax allele frequency of < 0.1% (popmax refers to the maximum frequency across continental populations in gnomAD, not including Ashkenazi Jewish, Finnish, or any other remaining samples) across the entire gnomAD dataset (v2 exomes, v2 genomes, and v3 genomes). Variants were also excluded if they had not passed gnomAD quality control or had been flagged as part of a problematic region (low complexity, decoy, and segmental duplication). Unique variants for the 100 individuals were obtained by filtering to variants with AC of 1 in gnomAD v2 exomes and no observations in v2 genomes or in the v3-non v2 subset.

The variant annotations previously applied by the Variant Effect Predictor (VEP; version 85) were used to assign each variant to one of three categories: pLoF, missense/inframe indel, or synonymous (McLaren et al., 2016). The VEP annotations were filtered to the most severe consequence for the canonical transcript. A variant was annotated as pLoF if the most severe consequence was one of: "splice_acceptor", "splice_donor_variant", "stop_gained", "frameshift_variant" and it was found to be high-confidence (indicated by LOFTEE) (Karczewski et al., 2020). The missense/indel annotation was applied if the most severe consequence was one of: "missense_variant", "inframe_insertion", "inframe_deletion". A synonymous annotation was used for a variant if the most severe consequence was "synonymous_variant". Analysis was performed using Hail (https://hail.is/). Variants were exported to a tsv for data visualization in R. The code used to select individuals and filter variants can be found on GitHub: https://github.com/broadinstitute/gnomad_review_hum_mut.

# SUPPLEMENTARY FIGURES AND TABLES

**Supplementary Table S1:** Mean count of very rare coding variants per population (allele frequency < 0.1%). Total unique refers to variants found in only that individual across the entire gnomAD dataset (v2 exome, v2 genome, v3).

| Population | pLoF | Missense/ indel | Synonymous | Total | Total unique |
|---|---|---|---|---|---|
| African/African American | 7 (±3) | 137 (±18) | 77 (±12) | 220 (±27) | 21 (±7) |
| Latino/Admixed American | 6 (±3) | 111 (±20) | 57 (±12) | 175 (±29) | 19 (±11) |
| East Asian | 7 (±3) | 127 (±21) | 66 (±13) | 201 (±32) | 35 (±11) |
| European (non-finnish) | 7 (±2) | 108 (±18) | 55 (±13) | 170 (±30) | 23 (±11) |
| South Asian | 7 (±3) | 147 (±22) | 80 (±14) | 235 (±33) | 38 (±14) |
| **Global** | **7 (±3)** | **126 (±25)** | **67 (±16)** | **200 (±39)** | **27 (±13)** |

(±) standard deviation

**Supplementary Table S2:** Comparison of mean count of unique coding variants per population utilizing only v2 exome dataset or the entire gnomAD dataset. The general decrease in the number of unique variants from mean 36 (± 16) to 27 (± 13) (p < 2.2e-16) across populations demonstrates the importance of increased sample size and more diverse representation of populations.

| Population | v2 exome only | Entire gnomAD | P-value |
|---|---|---|---|
| African/African American | 42 (±11) | 21 (±7) | 2.20E-16 |
| Latino/Admixed American | 26 (±18) | 19 (±11) | 3.49E-03 |
| East Asian | 42 (±13) | 35 (±11) | 7.85E-05 |
| European (non-Finnish) | 29 (±12) | 23 (±11) | 1.52E-04 |
| South Asian | 41 (±15) | 38 (±14) | 9.46E-02 |
| **Global** | **36 (±16)** | **27 (±13)** | **2.20E-16** |

P-value calculated using Student t-test.

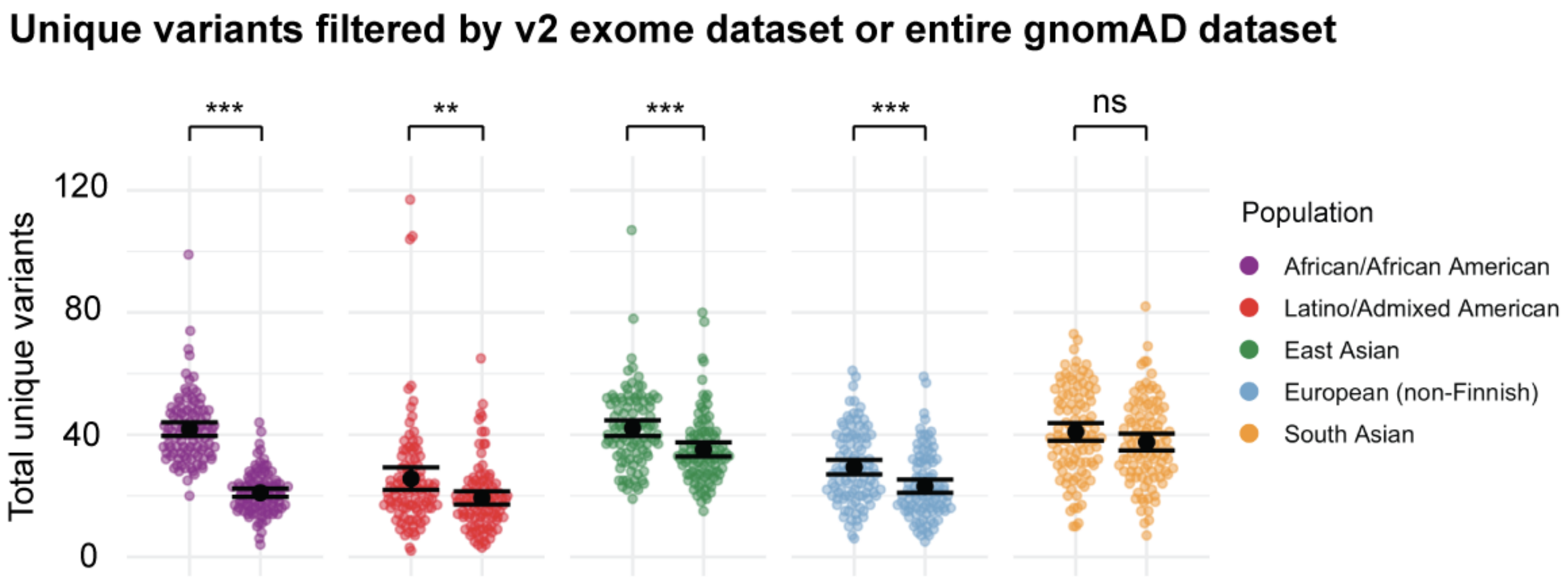

**Supplementary Figure S1:** Total unique variants in two sets of 100 randomly selected individuals filtered using v2 exomes (n = 125,748; left) versus the entire gnomAD dataset (n = 197,912; right). The increased sample size results in a decrease in the number of unique variants per individual in all populations. The increase from ~8,000 African/African American individuals in v2 exomes to ~30,000 individuals across the entire database results in a significant decrease in unique variants from mean 42 (± 11) to 21 (± 7) variants per individual in this population (purple). In comparison the minor increase in South Asian samples from ~15,000 to ~16,500 results in an non-significant decrease from mean 41 (± 15) to 38 (± 14) variants per individual (orange). Of note, the mean is influenced by the fact that these individuals are in gnomAD and likely from a community that has some representation in the database. An individual from a subpopulation or community that is not as well represented in gnomAD will likely have a higher number of unique very rare variants than seen in this analysis. Black bar represents the 95% confidence interval. ** $p < 0.01$, *** $p < 0.001$, ns: not significant with Student t-test. (±) Standard deviation.

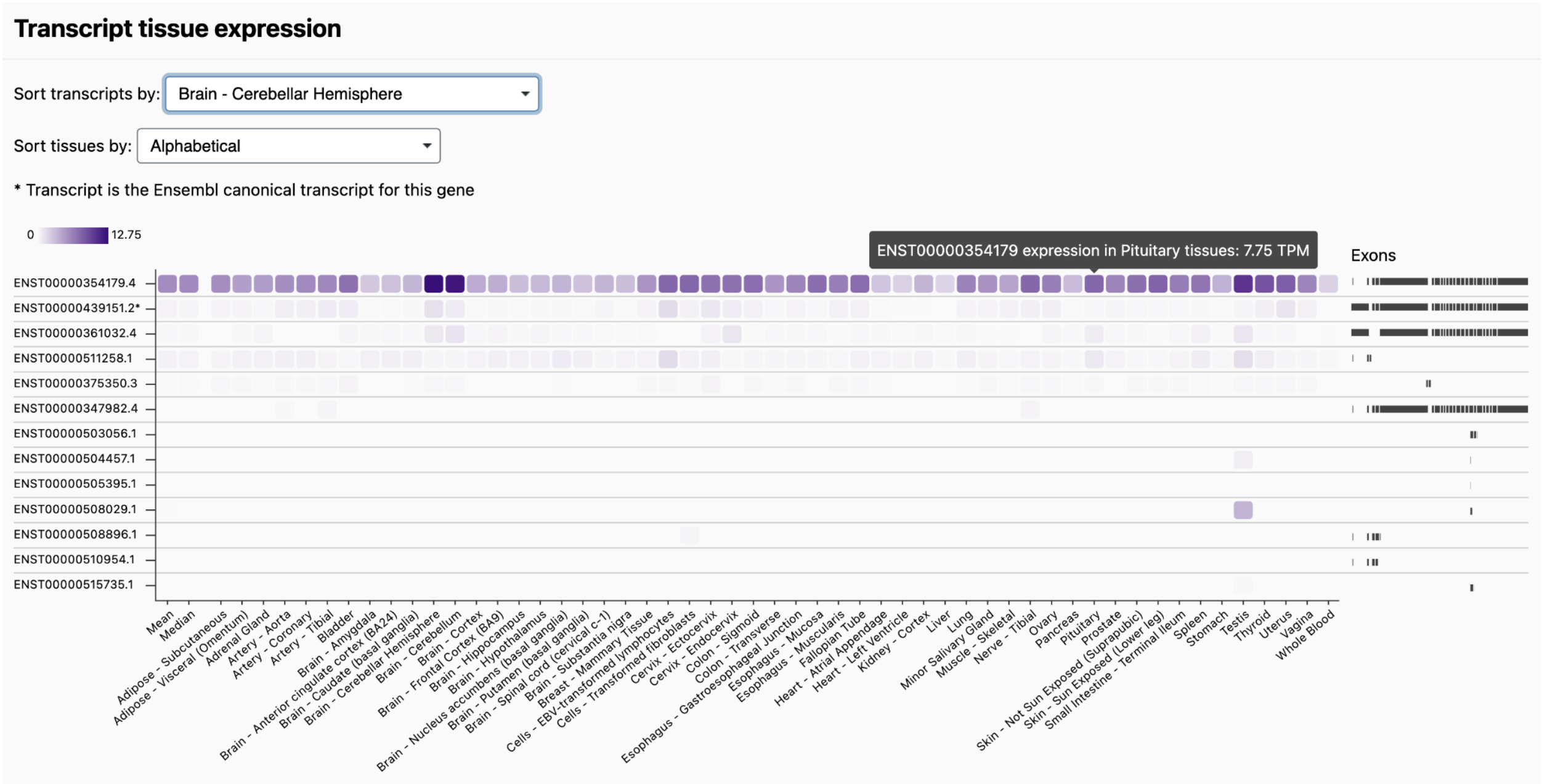

**Supplementary Figure S2:** Transcript tissue expression of *NSD1*. Navigate to the pop-up window by selecting “Show transcripts” (Figure 3:6) or “Show tissues” (Figure 3:11) on the gene page, and then “show transcript tissue expression” (Figure 3:8). Transcripts (y-axis) can be sorted by their expression in a specific tissue. In this example, sorting by Brain - Cerebellar Hemisphere lists ENST00000354179.4 first as the highest expressed transcript in that tissue. Tissues (x-axis) can be sorted alphabetically (this example) or by highest mean expression. A darker purple square indicates higher expression of a specific transcript in a specific tissue measured by transcripts per million (TPM). The exact value is available as a hover over (in this example highlighting the expression of ENST00000354179.4 in Pituitary tissue as 7.75 TPM). The highest expression is found in “Brain - Cerebellar Hemisphere” and “Brain - Cerebellum” (darkest purple).  *The canonical transcript.

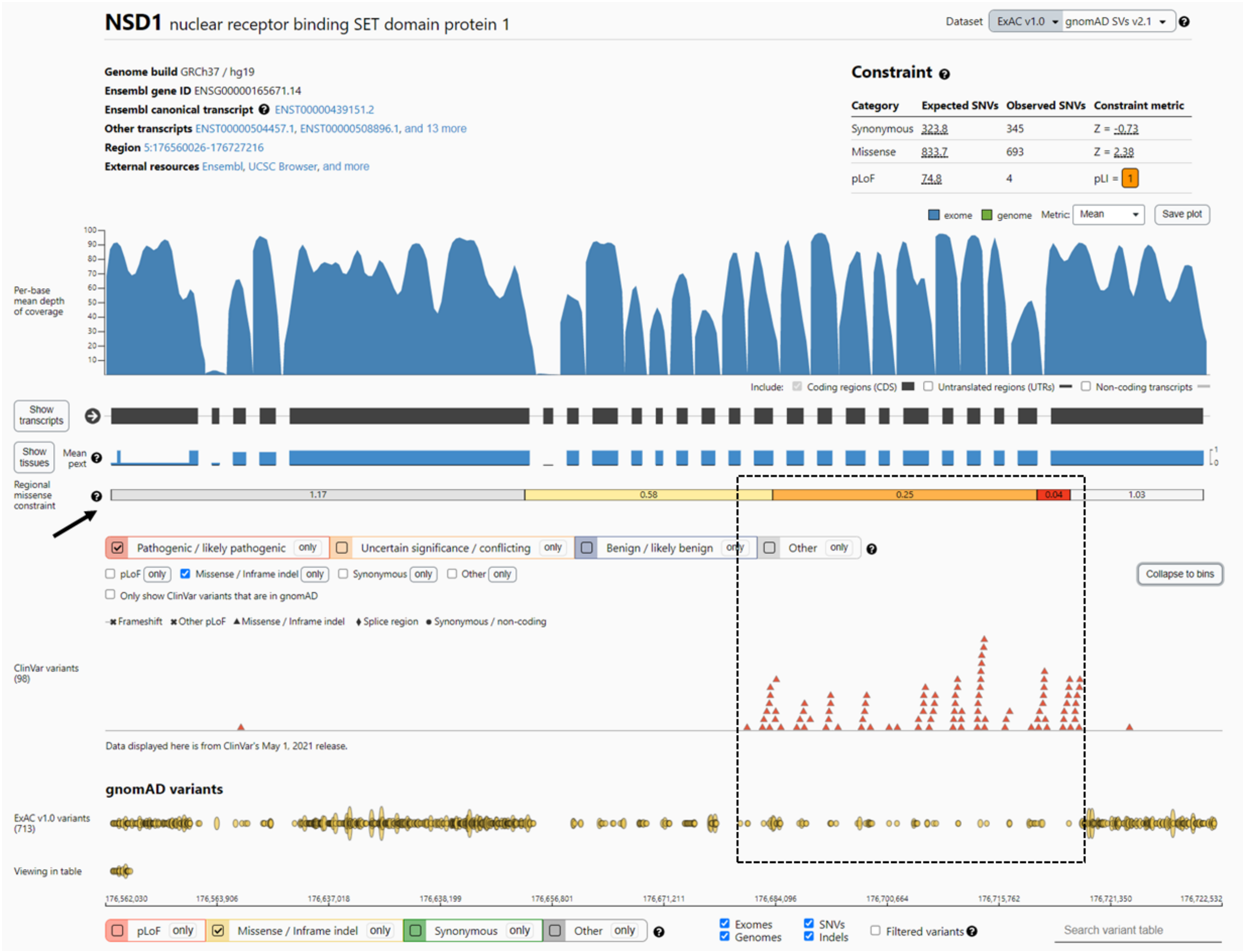

**Supplementary Figure S3:** ExAC dataset gene page for *NSD1*. The track is available under the pext track (black arrow) for genes with regional missense constraint. The numbers on the track refer to the proportion of observed versus expected missense variation seen in the region in gnomAD. In this example, missense constrained regions (red and orange) show clustering of pathogenic/likely pathogenic missense variants in ClinVar, while there is a depletion of missense variation in gnomAD for the same region (box).

| Variant ID | Source | HGVS Consequence | VEP Annotation | LoF Curation | Clinical Significance | Flags | Allele Count | Allele Number | Allele Frequency | Number of Homozygotes |
|---|---|---|---|---|---|---|---|---|---|---|
| 4-981738-G-T | E | c.299+1G>T | splice donor | Likely LoF | Pathogenic/Likely p... | | 1 | 178998 | 5.59e-6 | 0 |
| 4-985629-CG-C | E | p.Ile8LeufsTer31 † | frameshift | Not LoF | | pLoF flag | 3 | 127736 | 2.35e-5 | 0 |
| 4-985947-TCA-T | E | p.His8GlnfsTer54 † | frameshift | Not LoF | | pLoF flag | 8 | 128350 | 6.23e-5 | 0 |
| 4-985963-C-T | E | p.Gln13Ter † | stop gained | Not LoF | | pLoF flag | 1 | 128414 | 7.79e-6 | 0 |
| 4-985969-C-T | E | p.Gln15Ter † | stop gained | Not LoF | | pLoF flag | 2 | 128416 | 1.56e-5 | 0 |
| 4-994398-AG-A | E | p.Ser102ProfsTer6 | splice acceptor | Likely not LoF | | | 1 | 250076 | 4.00e-6 | 0 |
| 4-994400-G-A | E | p.Trp53Ter † | stop gained | LoF | | | 1 | 250112 | 4.00e-6 | 0 |
| 4-994417-GC-G | E | p.Leu107Ter | frameshift | LoF | | | 1 | 250322 | 3.99e-6 | 0 |
| 4-994486-G-C | E | c.385+1G>C | splice donor | LoF | Pathogenic | | 1 | 249514 | 4.01e-6 | 0 |
| 4-994668-A-G | E | c.386-2A>G | splice acceptor | LoF | Pathogenic | | 9 | 250194 | 3.60e-5 | 0 |
| 4-995255-G-A | E | c.494-1G>A | splice acceptor | Likely LoF | Pathogenic | | 1 | 250814 | 3.99e-6 | 0 |

**Supplementary Figure S4:** Variant table from *IDUA* showing the manual curation verdict (under LoF Curation), the LoF transcript effect estimator (LOFTEE) low-confidence (LC) or pLoF flag (under Flags), with further detail supplied on the variant page. Generally, manual curation was not necessary or performed if a variant was determined to be LC by LOFTEE. Manual curation verdicts are LoF, Likely LoF, Uncertain LoF, Likely not LoF, and Not LoF.

(a)

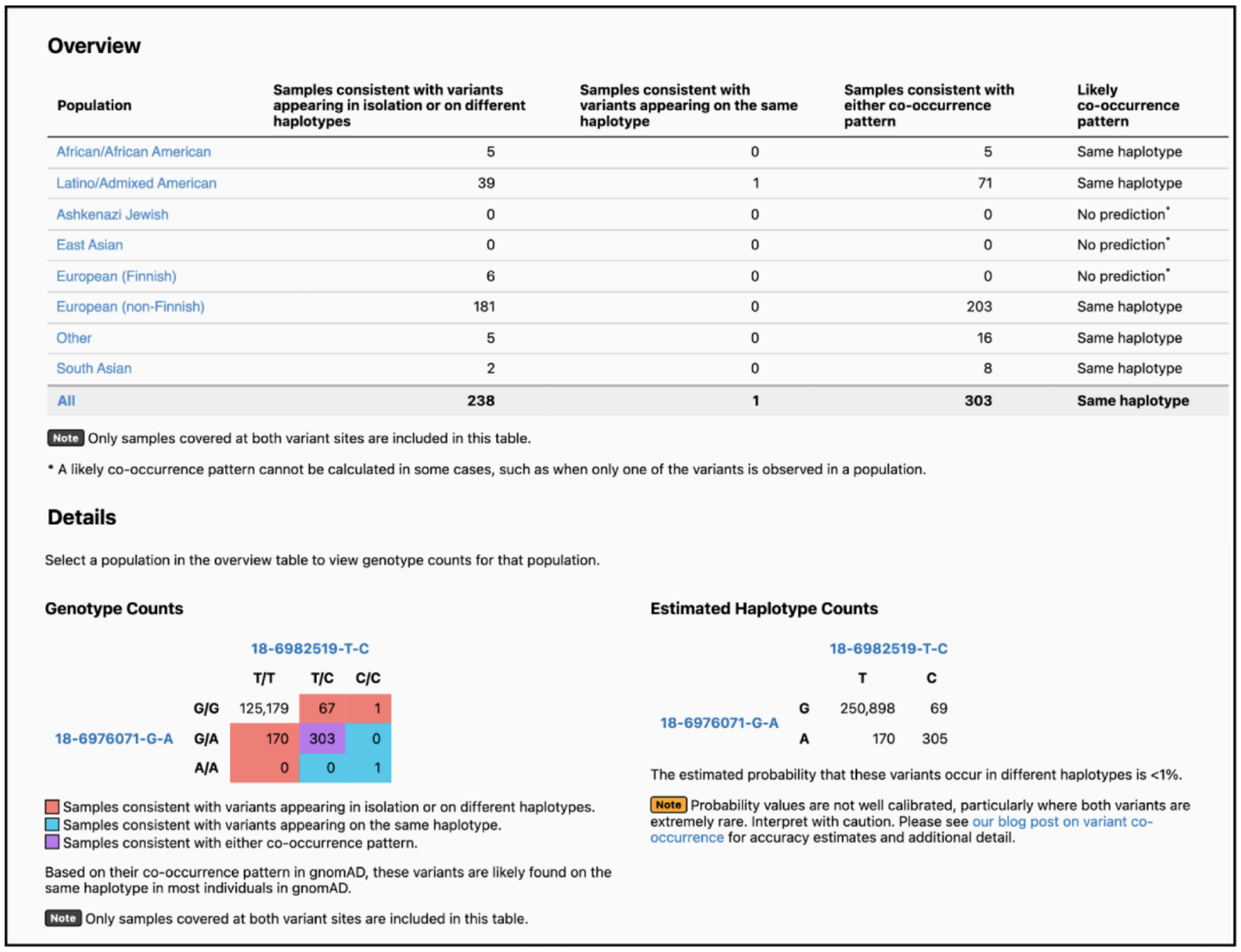

**Overview**

| Population | Samples consistent with variants appearing in isolation or on different haplotypes | Samples consistent with variants appearing on the same haplotype | Samples consistent with either co-occurrence pattern | Likely co-occurrence pattern |
|---|---|---|---|---|
| African/African American | 5 | 0 | 5 | Same haplotype |
| Latino/Admixed American | 39 | 1 | 71 | Same haplotype |
| Ashkenazi Jewish | 0 | 0 | 0 | No prediction* |
| East Asian | 0 | 0 | 0 | No prediction* |
| European (Finnish) | 6 | 0 | 0 | No prediction* |
| European (non-Finnish) | 181 | 0 | 203 | Same haplotype |
| Other | 5 | 0 | 16 | Same haplotype |
| South Asian | 2 | 0 | 8 | Same haplotype |
| **All** | **238** | **1** | **303** | **Same haplotype** |

Note Only samples covered at both variant sites are included in this table.

* A likely co-occurrence pattern cannot be calculated in some cases, such as when only one of the variants is observed in a population.

**Details**

Select a population in the overview table to view genotype counts for that population.

**Genotype Counts**

| 18-6976071-G-A \ 18-6982519-T-C | T/T | T/C | C/C |
|---|---|---|---|
| G/G | 125,179 | 67 | 1 |
| G/A | 170 | 303 | 0 |
| A/A | 0 | 0 | 1 |

Samples consistent with variants appearing in isolation or on different haplotypes.
Samples consistent with variants appearing on the same haplotype.
Samples consistent with either co-occurrence pattern.

Based on their co-occurrence pattern in gnomAD, these variants are likely found on the same haplotype in most individuals in gnomAD.

Note Only samples covered at both variant sites are included in this table.

**Estimated Haplotype Counts**

| 18-6976071-G-A \ 18-6982519-T-C | T | C |
|---|---|---|
| G | 250,898 | 69 |
| A | 170 | 305 |

The estimated probability that these variants occur in different haplotypes is <1%.

Note Probability values are not well calibrated, particularly where both variants are extremely rare. Interpret with caution. Please see our blog post on variant co-occurrence for accuracy estimates and additional detail.

(b)

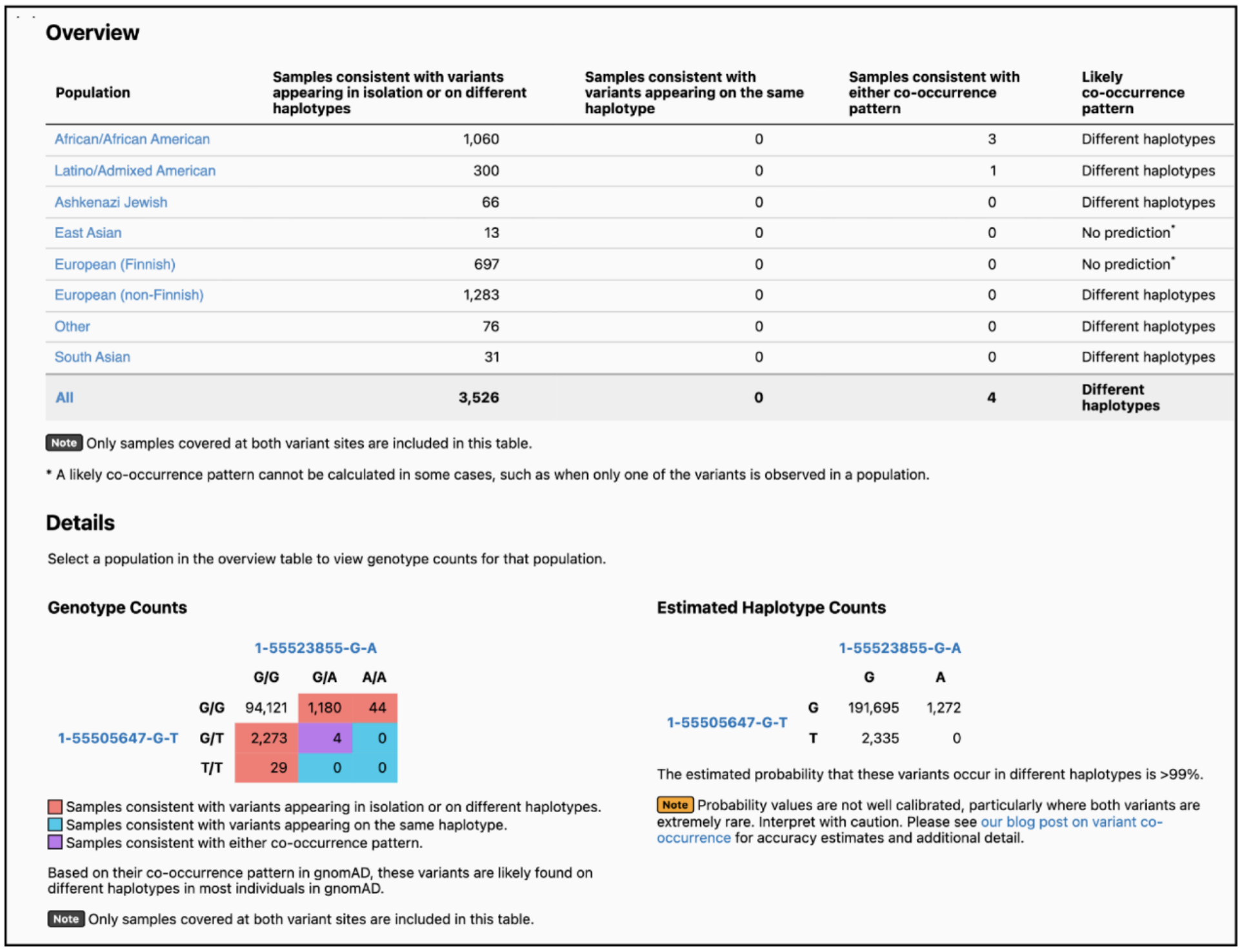

**Overview**

| Population | Samples consistent with variants appearing in isolation or on different haplotypes | Samples consistent with variants appearing on the same haplotype | Samples consistent with either co-occurrence pattern | Likely co-occurrence pattern |
|---|---|---|---|---|
| African/African American | 1,060 | 0 | 3 | Different haplotypes |
| Latino/Admixed American | 300 | 0 | 1 | Different haplotypes |
| Ashkenazi Jewish | 66 | 0 | 0 | Different haplotypes |
| East Asian | 13 | 0 | 0 | No prediction* |
| European (Finnish) | 697 | 0 | 0 | No prediction* |
| European (non-Finnish) | 1,283 | 0 | 0 | Different haplotypes |
| Other | 76 | 0 | 0 | Different haplotypes |
| South Asian | 31 | 0 | 0 | Different haplotypes |
| **All** | **3,526** | **0** | **4** | **Different haplotypes** |

Note Only samples covered at both variant sites are included in this table.

* A likely co-occurrence pattern cannot be calculated in some cases, such as when only one of the variants is observed in a population.

**Details**

Select a population in the overview table to view genotype counts for that population.

**Genotype Counts**

| 1-55505647-G-T \ 1-55523855-G-A | G/G | G/A | A/A |
|---|---|---|---|
| G/G | 94,121 | 1,180 | 44 |
| G/T | 2,273 | 4 | 0 |
| T/T | 29 | 0 | 0 |

Samples consistent with variants appearing in isolation or on different haplotypes.
Samples consistent with variants appearing on the same haplotype.
Samples consistent with either co-occurrence pattern.

Based on their co-occurrence pattern in gnomAD, these variants are likely found on different haplotypes in most individuals in gnomAD.

Note Only samples covered at both variant sites are included in this table.

**Estimated Haplotype Counts**

| 1-55505647-G-T \ 1-55523855-G-A | G | A |
|---|---|---|
| G | 191,695 | 1,272 |
| T | 2,335 | 0 |

The estimated probability that these variants occur in different haplotypes is >99%.

Note Probability values are not well calibrated, particularly where both variants are extremely rare. Interpret with caution. Please see our blog post on variant co-occurrence for accuracy estimates and additional detail.

**Supplementary Figure S5:** The variant co-occurrence feature shows the probability that two variants of interest are found on different haplotypes using a statistical phasing approach. (A) In the first example chr18:6976071G>A (NM_005559.4:c.6354C>T) and chr18:6982519T>C (NM_005559.4:c.5867A>G) in the *LAMA1* gene are investigated.

The variants are found in isolation in 68 and 170 individuals (red) and co-occur in 304 individuals (purple and blue). Given the high proportion of individuals where both variants are present, they are predicted to be on the same haplotype in the informative populations. East Asian and European (Finnish) are uninformative as there are no individuals harboring both variants in these populations. The variants must be on the same haplotype in the one individual (blue) who is homozygous for both variants. (B) In the second example chr1:55505647G>T (NM_174936.4:c.137G>T) and chr1:55523855G>A (NM_174936.4:c.1327G>A) in the *PCSK9* gene are investigated. They are reported to likely be on different haplotypes in all informative populations as they co-occur in only four individuals in gnomAD (purple). Thus, the presence of these four individuals in gnomAD suggest that compound heterozygosity for these two variants is not associated with a severe, early-onset, highly penetrant disorder.

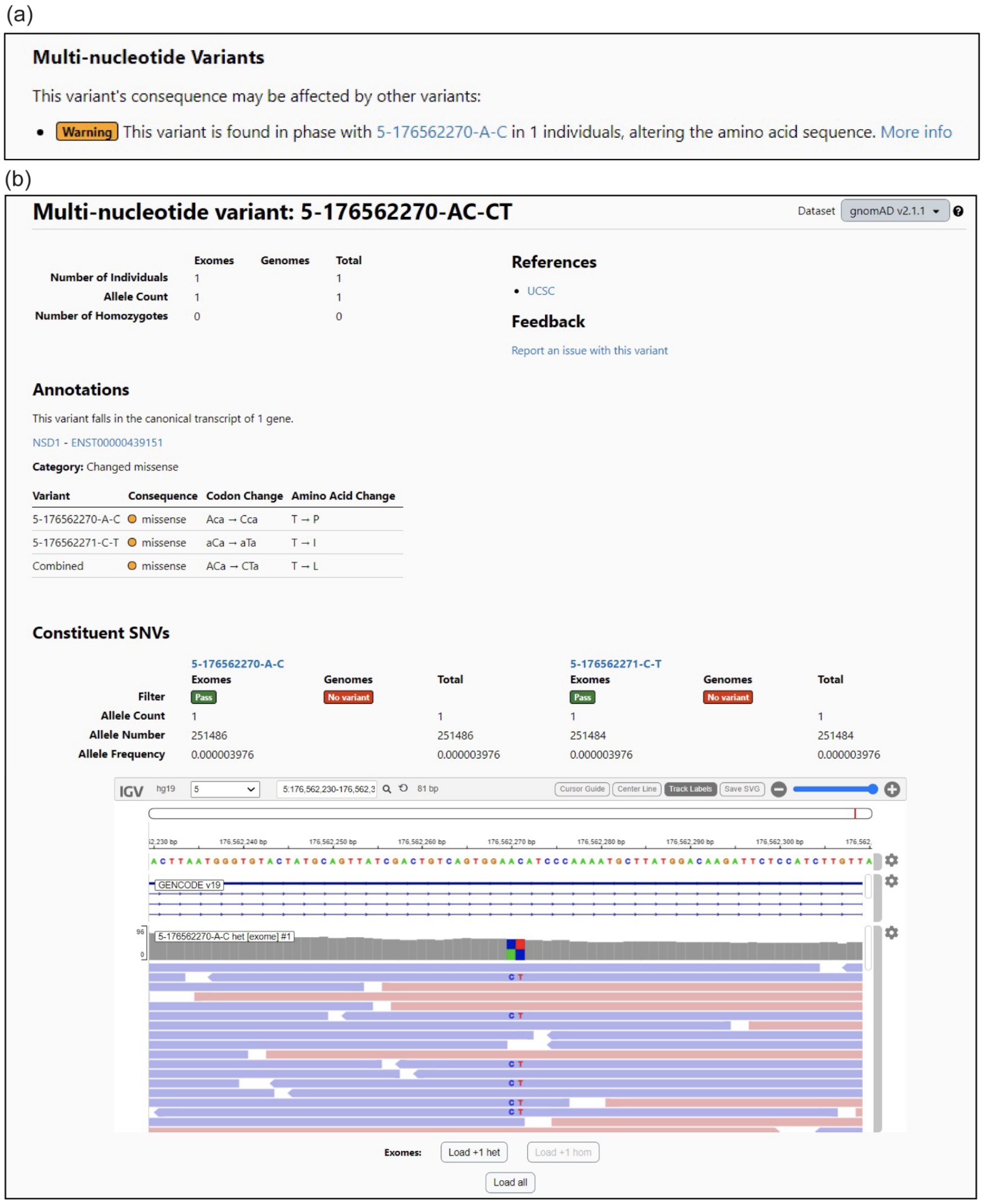

**Supplementary Figure S6:** Multi-nucleotide variant (MNV) in *MBD5*. (a) From the variant page for either of the individual variants, the warning indicates that the variant is part of an MNV. (b) The header for the MNV page displays frequency information for the MNV. The predicted variant type of each separate variant, as well as for the combined variant type is noted in the table (in this case a stop-gained and missense variant within this codon in combination produces a missense variant). The variant links navigate directly to the page for each variant.

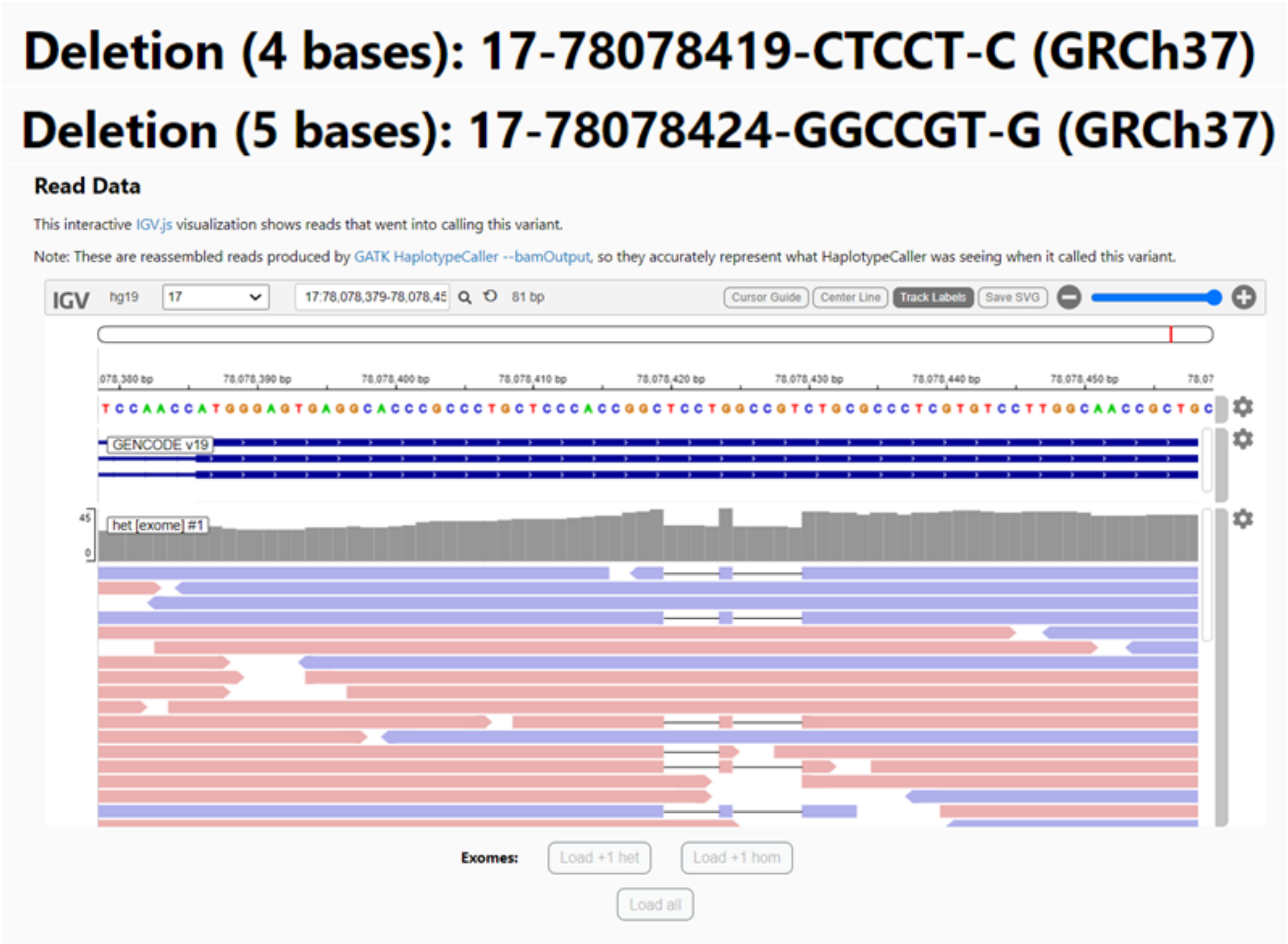

**Supplementary Figure S7:** Frame-restoring indel in the *GAA* gene. While each variant interpreted independently would result in a frameshift, in combination these result in a single 9 base-pair (3 amino acid) in-frame deletion. The effect of the variant is not LoF and should be interpreted as an in-frame indel. Indels (either as SNV-indel variants or indel-indel variants) are not currently annotated in the gnomAD browser as part of MNVs.

(a)

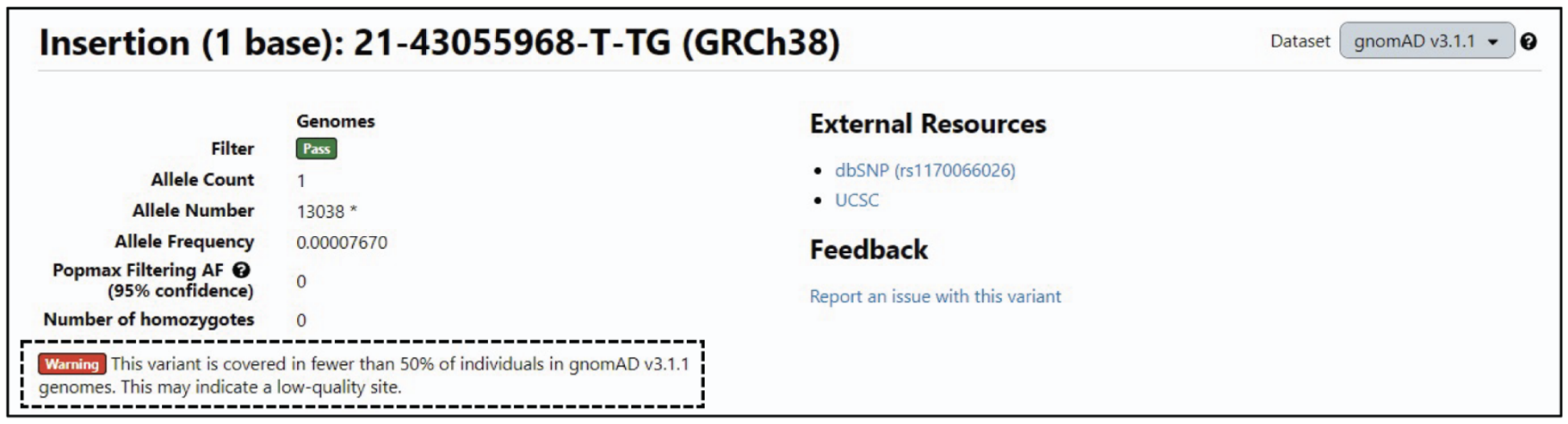

(b)

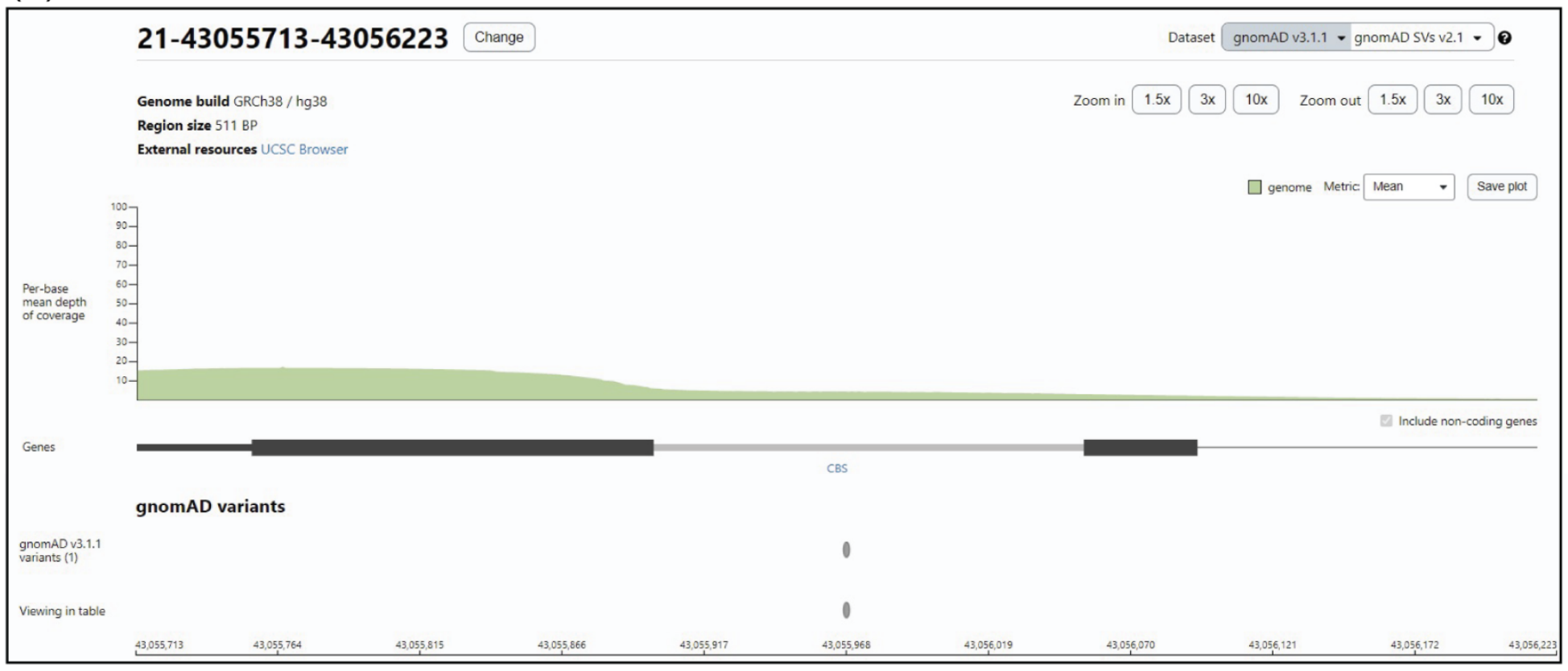

(c)

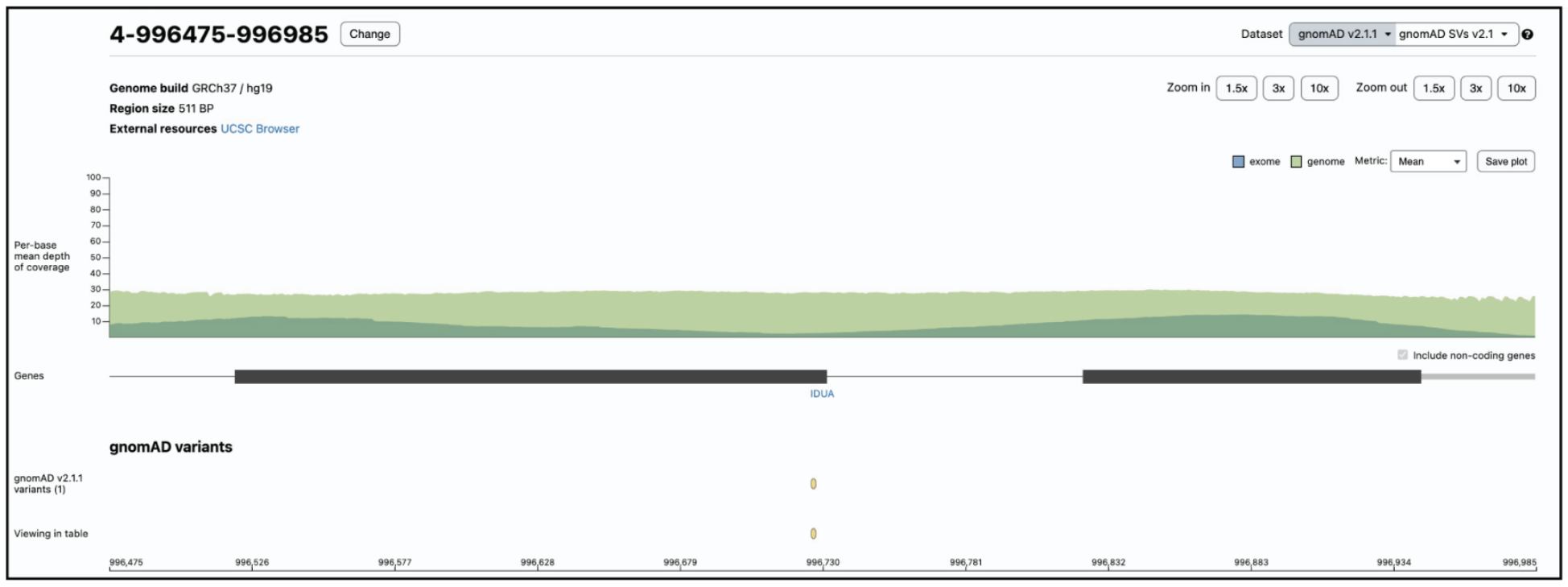

**Supplementary Figure S8:** Low coverage warning and caveats. (a) Top panel displays a warning (dashed box) that the variant in the *CBS* gene is covered in fewer than 50% of individuals. (b) Regional view of the *CBS* gene (searching for an interval in the search box) displaying low coverage in genomes (gnomAD v3) for the region of the highlighted variant (gray oval). (c) Regional view of *IDUA* gene displaying low coverage in exomes (blue), but not genome (green) in gnomAD v2 for the region of the highlighted variant (yellow oval). Allele frequencies in the region are likely reliable in the *IUDA* gene from the genome (but not exome) data (c) but not reliable in *CBS* gene from either genome or exome data (b).

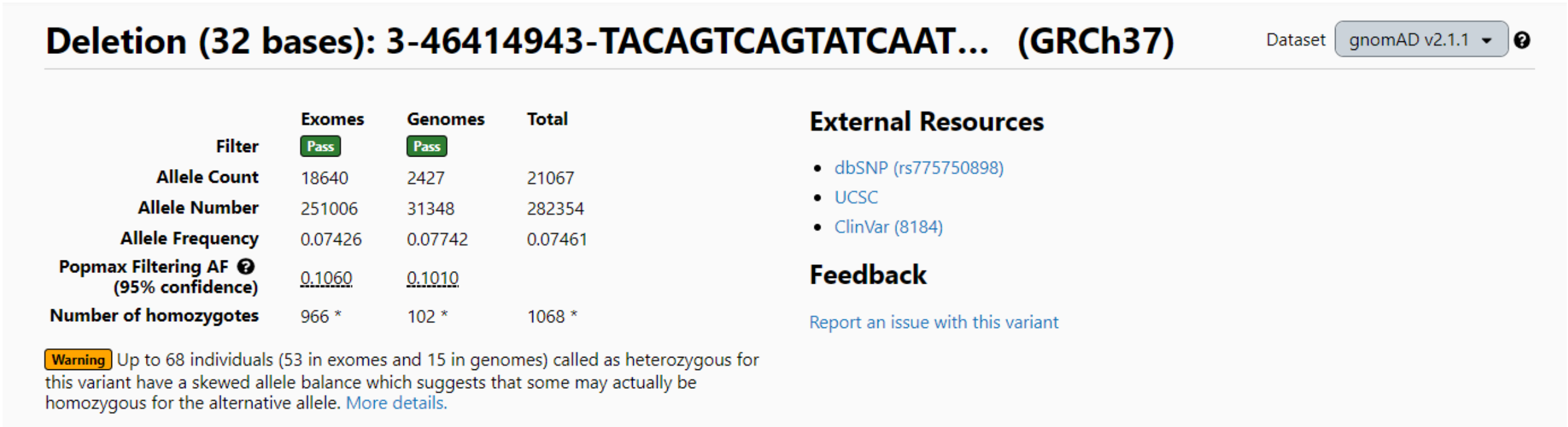

**Supplementary Figure S9:** Flag for variant with a skewed allele balance for homozygous variants. Heterozygous variants with inflated alt-reads are likely homozygous variants called heterozygotes due to contamination that affects the variant calling likelihood models. This depletion of homozygous calls can incorrectly deflate the allele frequencies (Karczewski et al., 2019).

## SUPPLEMENTARY REFERENCES

## GENOME AGGREGATION DATABASE CONSORTIUM AUTHORS

Maria Abreu[1], Carlos A. Aguilar Salinas[2], Tariq Ahmad[3], Christine M. Albert[4,5], Jessica Alföldi[6,7], Diego Ardissino[8], Irina M. Armean[6,7,9], Gil Atzmon[10,11], Eric Banks[12], John Barnard[13], Samantha M. Baxter[6], Laurent Beaugerie[14], Emelia J. Benjamin[15,16,17], David Benjamin[12], Louis Bergelson[12], Michael Boehnke[18], Lori L. Bonnycastle[19], Erwin P. Bottinger[20], Donald W. Bowden[21,22,23], Matthew J. Bown[24,25], Steven Brant[26], Sarah E. Calvo[6,27], Hannia Campos[28,29], John C. Chambers[30,31,32], Juliana C. Chan[33], Katherine R. Chao[6], Sinéad Chapman[6,7,34], Daniel Chasman[4,35], Rex Chisholm[36], Judy Cho[20], Rajiv Chowdhury[37], Mina K. Chung[38], Wendy Chung[39,40,41], Kristian Cibulskis[12], Bruce Cohen[35,42], Ryan L. Collins[6,27,43], Kristen M. Connolly[44], Adolfo Correa[45], Miguel Covarrubias[12], Beryl Cummings[6,43], Dana Dabelea[46], Mark J. Daly[6,7,47], John Danesh[37], Dawood Darbar[48], Joshua Denny[49], Stacey Donnelly[6], Ravindranath Duggirala[50], Josée Dupuis[51,52], Patrick T. Ellinor[6,53], Roberto Elosua[54,55,56], James Emery[12], Eleina England[6], Jeanette Erdmann[57,58,59], Tõnu Esko[6,60], Emily Evangelista[6], Yossi Farjoun[12], Diane Fatkin[61,62,63], Steven Ferriera[64], Jose Florez[35,65,66], Laurent Francioli[6,7], Andre Franke[67], Martti Färkkilä[68], Stacey Gabriel[64], Kiran Garimella[12], Laura D. Gauthier[12], Jeff Gentry[12], Gad Getz[35,69,70], David C. Glahn[71,72], Benjamin Glaser[73], Stephen J. Glatt[74], David Goldstein[75,76], Clicerio Gonzalez[77], Julia Goodrich[6], Leif Groop[78,79], Sanna Gudmundsson[6,7,80], Namrata Gupta[6,64], Andrea Haessly[12], Christopher Haiman[81], Ira Hall[82], Craig Hanis[83], Matthew Harms[84,85], Mikko Hiltunen[86], Matti M. Holi[87], Christina M. Hultman[88,89], Chaim Jalas[90], Thibault Jeandet[12], Mikko Kallela[91], Diane Kaplan[12], Jaakko Kaprio[79], Konrad J. Karczewski[6,7,34], Sekar Kathiresan[27,35,92], Eimear Kenny[89,93], Bong-Jo Kim[94], Young Jin Kim[94], George Kirov[95], Jaspal Kooner[31,96,97], Seppo Koskinen[98], Harlan M. Krumholz[99], Subra Kugathasan[100], Soo Heon Kwak[101], Markku Laakso[102,103], Nicole Lake[104], Trevyn Langsford[12], Kristen M. Laricchia[6,7], Terho Lehtimäki[105], Monkol Lek[104], Emily Lipscomb[6], Christopher Llanwarne[12], Ruth J.F. Loos[20,106], Steven A. Lubitz[6,53], Teresa Tusie Luna[107,108], Ronald C.W. Ma[33,109,110], Daniel G. MacArthur[6,111,112], Gregory M. Marcus[113], Jaume Marrugat[55,114], Kari M. Mattila[105], Steven McCarroll[34,115], Mark I. McCarthy[116,117,118], Jacob McCauley[119,120], Dermot McGovern[121], Ruth McPherson[122], James B. Meigs[6,35,123], Olle Melander[124],

Andres Metspalu[125], Deborah Meyers[126], Eric V. Minikel[6], Braxton Mitchell[127], Vamsi K. Mootha[6,128], Ruchi Munshi[12], Aliya Naheed[129], Saman Nazarian[130,131], Benjamin M. Neale[6,7], Peter M. Nilsson[132], Sam Novod[12], Anne H. O'Donnell-Luria[6,7,80], Michael C. O'Donovan[95], Yukinori Okada[133,134,135], Dost Ongur[35,42], Lorena Orozco[136], Michael J. Owen[95], Colin Palmer[137], Nicholette D. Palmer[138], Aarno Palotie[7,34,79], Kyong Soo Park[101,139], Carlos Pato[140], Nikelle Petrillo[12], William Phu[6,80], Timothy Poterba[6,7,34], Ann E. Pulver[141], Dan Rader[130,142], Nazneen Rahman[143], Heidi Rehm[6,27], Alex Reiner[144,145], Anne M. Remes[146], Dan Rhodes[6], Stephen Rich[147,148], John D. Rioux[149,150], Samuli Ripatti[79,151,152], David Roazen[12], Dan M. Roden[153,154], Jerome I. Rotter[155], Valentin Ruano-Rubio[12], Nareh Sahakian[12], Danish Saleheen[156,157,158], Veikko Salomaa[159], Andrea Saltzman[6], Nilesh J. Samani[24,25], Kaitlin E. Samocha[160], Jeremiah Scharf[6,27,34], Molly Schleicher[6], Heribert Schunkert[161,162], Sebastian Schönherr[163], Eleanor Seaby[6], Cotton Seed[7,34], Svati H. Shah[164], Megan Shand[12], Moore B. Shoemaker[165], Tai Shyong[166,167], Edwin K. Silverman[168,169], Moriel Singer-Berk[6], Pamela Sklar[170,171,172], J. Gustav Smith[152,173,174], Jonathan T. Smith[12], Hilkka Soininen[175], Harry Sokol[176,177,178], Matthew Solomonson[6,7], Rachel G. Son[6], Jose Soto[12], Tim Spector[179], Christine Stevens[6,7,34], Nathan Stitziel[82,180], Patrick F. Sullivan[88,181], Jaana Suvisaari[159], E. Shyong Tai[182,183,184], Michael E. Talkowski[6,27,34], Yekaterina Tarasova[6], Kent D. Taylor[155], Yik Ying Teo[182,185,186], Grace Tiao[6,7], Kathleen Tibbetts[12], Charlotte Tolonen[12], Ming Tsuang[187,188], Tiinamaija Tuomi[79,189,190], Dan Turner[191], Teresa Tusie-Luna[192,193], Erkki Vartiainen[194], Marquis Vawter[195], Christopher Vittal[6,7], Gordon Wade[12], Arcturus Wang[6,7,34], Qingbo Wang[6,133], James S. Ware[6,196,197], Hugh Watkins[198], Nicholas A. Watts[6,7], Rinse K. Weersma[199], Ben Weisburd[12], Maija Wessman[79,200], Nicola Whiffin[6,201,202], Michael W. Wilson[6,7], James G. Wilson[203], Ramnik J. Xavier[204,205]

[1]University of Miami Miller School of Medicine, Gastroenterology, Miami, USA

[2]Unidad de Investigacion de Enfermedades Metabolicas, Instituto Nacional de Ciencias Medicas y Nutricion, Mexico City, Mexico

[3]Peninsula College of Medicine and Dentistry, Exeter, UK

[4]Division of Preventive Medicine, Brigham and Women's Hospital, Boston, MA, USA

[5]Division of Cardiovascular Medicine, Brigham and Women's Hospital and Harvard Medical School, Boston, MA, USA
[6]Program in Medical and Population Genetics, Broad Institute of MIT and Harvard, Cambridge, MA, USA
[7]Analytic and Translational Genetics Unit, Massachusetts General Hospital, Boston, MA, USA
[8]Department of Cardiology University Hospital, Parma, Italy
[9]European Molecular Biology Laboratory, European Bioinformatics Institute, Wellcome Genome Campus, Hinxton, Cambridge, UK
[10]Department of Biology Faculty of Natural Sciences, University of Haifa, Haifa, Israel
[11]Departments of Medicine and Genetics, Albert Einstein College of Medicine, Bronx, NY, USA
[12]Data Science Platform, Broad Institute of MIT and Harvard, Cambridge, MA, USA
[13]Department of Quantitative Health Sciences, Lerner Research Institute Cleveland Clinic, Cleveland, OH, USA
[14]Sorbonne Université, APHP, Gastroenterology Department Saint Antoine Hospital, Paris, France
[15]NHLBI and Boston University's Framingham Heart Study, Framingham, MA, USA
[16]Department of Medicine, Boston University School of Medicine, Boston, MA, USA
[17]Department of Epidemiology, Boston University School of Public Health, Boston, MA, USA
[18]Department of Biostatistics and Center for Statistical Genetics, University of Michigan, Ann Arbor, MI, USA
[19]National Human Genome Research Institute, National Institutes of Health Bethesda, MD, USA
[20]The Charles Bronfman Institute for Personalized Medicine, Icahn School of Medicine at Mount Sinai, New York, NY, USA
[21]Department of Biochemistry, Wake Forest School of Medicine, Winston-Salem, NC, USA
[22]Center for Genomics and Personalized Medicine Research, Wake Forest School of Medicine, Winston-Salem, NC, USA

[23]Center for Diabetes Research, Wake Forest School of Medicine, Winston-Salem, NC, USA
[24]Department of Cardiovascular Sciences, University of Leicester, Leicester, UK
[25]NIHR Leicester Biomedical Research Centre, Glenfield Hospital, Leicester, UK
[26]John Hopkins Bloomberg School of Public Health, Baltimore, MD, USA
[27]Center for Genomic Medicine, Massachusetts General Hospital, Boston, MA, USA
[28]Harvard School of Public Health, Boston, MA, USA
[29]Central American Population Center, San Pedro, Costa Rica
[30]Department of Epidemiology and Biostatistics, Imperial College London, London, UK
[31]Department of Cardiology, Ealing Hospital, NHS Trust, Southall, UK
[32]Imperial College, Healthcare NHS Trust Imperial College London, London, UK
[33]Department of Medicine and Therapeutics, The Chinese University of Hong Kong, Hong Kong, China
[34]Stanley Center for Psychiatric Research, Broad Institute of MIT and Harvard, Cambridge, MA, USA
[35]Department of Medicine, Harvard Medical School, Boston, MA, USA
[36]Northwestern University, Evanston, IL, USA
[37]University of Cambridge, Cambridge, England
[38]Departments of Cardiovascular, Medicine Cellular and Molecular Medicine Molecular Cardiology, Quantitative Health Sciences, Cleveland Clinic, Cleveland, OH, USA
[39]Department of Pediatrics, Columbia University Irving Medical Center, New York, NY, USA
[40]Herbert Irving Comprehensive Cancer Center, Columbia University Medical Center, New York, NY, USA
[41]Department of Medicine, Columbia University Medical Center, New York, NY, USA
[42]McLean Hospital, Belmont, MA, USA
[43]Division of Medical Sciences, Harvard Medical School, Boston, MA, USA
[44]Genomics Platform, Broad Institute of MIT and Harvard, Cambridge, MA, USA
[45]Department of Medicine, University of Mississippi Medical Center, Jackson, MI, USA
[46]Department of Epidemiology Colorado School of Public Health Aurora, CO, USA
[47]Institute for Molecular Medicine Finland, (FIMM) Helsinki, Finland

[48]Department of Medicine and Pharmacology, University of Illinois at Chicago, Chicago, IL, USA
[49]Vanderbilt University Medical Center, Nashville, TN, USA
[50]Department of Genetics, Texas Biomedical Research Institute, San Antonio, TX, USA
[51]Department of Biostatistics, Boston University School of Public Health, Boston, MA, USA
[52]National Heart Lung and Blood Institute's Framingham Heart Study, Framingham, MA, USA
[53]Cardiac Arrhythmia Service and Cardiovascular Research Center, Massachusetts General Hospital, Boston, MA, USA
[54]Cardiovascular Epidemiology and Genetics Hospital del Mar Medical Research Institute, (IMIM) Barcelona Catalonia, Spain
[55]CIBER CV Barcelona, Catalonia, Spain
[56]Departament of Medicine, Medical School University of Vic-Central, University of Catalonia, Vic Catalonia, Spain
[57]Institute for Cardiogenetics, University of Lübeck, Lübeck, Germany
[58]German Research Centre for Cardiovascular Research, Hamburg/Lübeck/Kiel, Lübeck, Germany
[59]University Heart Center Lübeck, Lübeck, Germany
[60]Estonian Genome Center, Institute of Genomics University of Tartu, Tartu, Estonia
[61]Victor Chang Cardiac Research Institute, Darlinghurst, NSW, Australia
[62]Faculty of Medicine, UNSW Sydney, Kensington, NSW, Australia
[63]Cardiology Department, St Vincent's Hospital, Darlinghurst, NSW, Australia
[64]Broad Genomics, Broad Institute of MIT and Harvard, Cambridge, MA, USA
[65]Diabetes Unit and Center for Genomic Medicine, Massachusetts General Hospital, Boston, MA, USA
[66]Programs in Metabolism and Medical & Population Genetics, Broad Institute of MIT and Harvard, Cambridge, MA, USA
[67]Institute of Clinical Molecular Biology, (IKMB) Christian-Albrechts-University of Kiel, Kiel, Germany

[68]Helsinki University and Helsinki University Hospital Clinic of Gastroenterology, Helsinki, Finland
[69]Bioinformatics Program MGH Cancer Center and Department of Pathology, Boston, MA, USA
[70]Cancer Genome Computational Analysis, Broad Institute of MIT and Harvard, Cambridge, MA, USA
[71]Department of Psychiatry and Behavioral Sciences, Boston Children's Hospitaland Harvard Medical School, Boston, MA, USA
[72]Harvard Medical School Teaching Hospital, Boston, MA, USA
[73]Department of Endocrinology and Metabolism, Hadassah Medical Center and Faculty of Medicine, Hebrew University of Jerusalem, Israel
[74]Department of Psychiatry and Behavioral Sciences, SUNY Upstate Medical University, Syracuse, NY, USA
[75]Institute for Genomic Medicine, Columbia University Medical Center Hammer Health Sciences, New York, NY, USA
[76]Department of Genetics & Development Columbia University Medical Center, Hammer Health Sciences, New York, NY, USA
[77]Centro de Investigacion en Salud Poblacional, Instituto Nacional de Salud Publica, Mexico
[78]Lund University Sweden, Sweden
[79]Institute for Molecular Medicine Finland, (FIMM) HiLIFE University of Helsinki, Helsinki, Finland
[80]Division of Genetics and Genomics, Boston Children's Hospital, Boston, MA, USA
[81]Lund University Diabetes Centre, Malmö, Skåne County, Sweden
[82]Washington School of Medicine, St Louis, MI, USA
[83]Human Genetics Center University of Texas Health Science Center at Houston, Houston, TX, USA
[84]Department of Neurology Columbia University, New York City, NY, USA
[85]Institute of Genomic Medicine, Columbia University, New York City, NY, USA
[86]Institute of Biomedicine, University of Eastern Finland, Kuopio, Finland

[87]Department of Psychiatry, Helsinki University Central Hospital Lapinlahdentie, Helsinki, Finland
[88]Department of Medical Epidemiology and Biostatistics, Karolinska Institutet, Stockholm, Sweden
[89]Icahn School of Medicine at Mount Sinai, New York, NY, USA
[90]Bonei Olam, Center for Rare Jewish Genetic Diseases, Brooklyn, NY, USA
[91]Department of Neurology, Helsinki University, Central Hospital, Helsinki, Finland
[92]Cardiovascular Disease Initiative and Program in Medical and Population Genetics, Broad Institute of MIT and Harvard, Cambridge, MA, USA
[93]Charles Bronfman Institute for Personalized Medicine, New York, NY, USA
[94]Division of Genome Science, Department of Precision Medicine, National Institute of Health, Republic of Korea
[95]MRC Centre for Neuropsychiatric Genetics & Genomics, Cardiff University School of Medicine, Cardiff, Wales
[96]Imperial College, Healthcare NHS Trust, London, UK
[97]National Heart and Lung Institute Cardiovascular Sciences, Hammersmith Campus, Imperial College London, London, UK
[98]Department of Health THL-National Institute for Health and Welfare, Helsinki, Finland
[99]Section of Cardiovascular Medicine, Department of Internal Medicine, Yale School of Medicine, Center for Outcomes Research and Evaluation Yale-New Haven Hospital, New Haven, CT, USA
[100]Division of Pediatric Gastroenterology, Emory University School of Medicine, Atlanta, GA, USA
[101]Department of Internal Medicine, Seoul National University Hospital, Seoul, Republic of Korea
[102]The University of Eastern Finland, Institute of Clinical Medicine, Kuopio, Finland
[103]Kuopio University Hospital, Kuopio, Finland
[104]Department of Genetics, Yale School of Medicine, New Haven, CT, USA
[105]Department of Clinical Chemistry Fimlab Laboratories and Finnish Cardiovascular Research Center-Tampere Faculty of Medicine and Health Technology, Tampere University, Finland

[106]The Mindich Child Health and Development, Institute Icahn School of Medicine at Mount Sinai, New York, NY, USA
[107]National Autonomous University of Mexico, Mexico City, Mexico
[108]Salvador Zubirán National Institute of Health Sciences and Nutrition, Mexico City, Mexico
[109]Li Ka Shing Institute of Health Sciences, The Chinese University of Hong Kong, Hong Kong, China
[110]Hong Kong Institute of Diabetes and Obesity, The Chinese University of Hong Kong, Hong Kong, China
[111]Centre for Population Genomics, Garvan Institute of Medical Research and UNSW Sydney, Sydney, Australia
[112]Centre for Population Genomics, Murdoch Children's Research Institute, Melbourne, Australia
[113]University of California San Francisco Parnassus Campus, San Francisco, CA, USA
[114]Cardiovascular Research REGICOR Group, Hospital del Mar Medical Research Institute, (IMIM) Barcelona, Catalonia, Spain
[115]Department of Genetics, Harvard Medical School, Boston, MA, USA
[116]Oxford Centre for Diabetes, Endocrinology and Metabolism, University of Oxford, Churchill Hospital Old Road Headington, Oxford, OX, LJ, UK
[117]Welcome Centre for Human Genetics, University of Oxford, Oxford, OX, BN, UK
[118]Oxford NIHR Biomedical Research Centre, Oxford University Hospitals, NHS Foundation Trust, John Radcliffe Hospital, Oxford, OX, DU, UK
[119]John P. Hussman Institute for Human Genomics, Leonard M. Miller School of Medicine, University of Miami, Miami, FL, USA
[120]The Dr. John T. Macdonald Foundation Department of Human Genetics, Leonard M. Miller School of Medicine, University of Miami, Miami, FL, USA
[121]F. Widjaja Foundation Inflammatory Bowel and Immunobiology Research Institute Cedars-Sinai Medical Center, Los Angeles, CA, USA
[122]Atherogenomics Laboratory University of Ottawa, Heart Institute, Ottawa, Canada
[123]Division of General Internal Medicine, Massachusetts General Hospital, Boston, MA, USA

[124]Department of Clinical Sciences University, Hospital Malmo Clinical Research Center, Lund University, Malmö, Sweden
[125]Estonian Genome Center, Institute of Genomics, University of Tartu, Tartu, Estonia
[126]University of Arizona Health Science, Tuscon, AZ, USA
[127]University of Maryland School of Medicine, Baltimore, MD, USA
[128]Howard Hughes Medical Institute and Department of Molecular Biology, Massachusetts General Hospital, Boston, MA, USA
[129]International Centre for Diarrhoeal Disease Research, Bangladesh
[130]Perelman School of Medicine, University of Pennsylvania, Philadelphia, PA, USA
[131]Johns Hopkins Bloomberg School of Public Health, Baltimore, MD, USA
[132]Lund University, Dept. Clinical Sciences, Skåne University Hospital, Malmö, Sweden
[133]Department of Statistical Genetics, Osaka University Graduate School of Medicine, Suita, Japan
[134]Laboratory of Statistical Immunology, Immunology Frontier Research Center (WPI-IFReC), Osaka University, Suita, Japan
[135]Integrated Frontier Research for Medical Science Division, Institute for Open and Transdisciplinary Research Initiatives, Osaka University, Suita, Japan
[136]Instituto Nacional de Medicina Genómica, (INMEGEN) Mexico City, Mexico
[137]Medical Research Institute, Ninewells Hospital and Medical School University of Dundee, Dundee, UK
[138]Wake Forest School of Medicine, Winston-Salem, NC, USA
[139]Department of Molecular Medicine and Biopharmaceutical Sciences, Graduate School of Convergence Science and Technology, Seoul National University, Seoul, Republic of Korea
[140]Department of Psychiatry Keck School of Medicine at the University of Southern California, Los Angeles, CA, USA
[141]Department of Psychiatry and Behavioral Sciences, Johns Hopkins University School of Medicine, Baltimore, MD, USA
[142]Children's Hospital of Philadelphia, Philadelphia, PA, USA
[143]Division of Genetics and Epidemiology, Institute of Cancer Research, London, SM, NG

[144]University of Washington, Seattle, WA, USA
[145]Fred Hutchinson Cancer Research Center, Seattle, WA, USA
[146]Medical Research Center, Oulu University Hospital, Oulu Finland and Research Unit of Clinical Neuroscience Neurology University of Oulu, Oulu, Finland
[147]Center for Public Health Genomics, University of Virginia, Charlottesville, VA, USA
[148]Department of Public Health Sciences, University of Virginia, Charlottesville, VA, USA
[149]Research Center Montreal Heart Institute, Montreal, Quebec, Canada
[150]Department of Medicine, Faculty of Medicine Université de Montréal, Québec, Canada
[151]Department of Public Health Faculty of Medicine, University of Helsinki, Helsinki, Finland
[152]Broad Institute of MIT and Harvard, Cambridge, MA, USA
[153]Department of Biomedical Informatics Vanderbilt, University Medical Center, Nashville, TN, USA
[154]Department of Medicine, Vanderbilt University Medical Center, Nashville, TN, USA
[155]The Institute for Translational Genomics and Population Sciences, Department of Pediatrics, The Lundquist Institute for Biomedical Innovation at Harbor-UCLA Medical Center, Torrance, CA, USA
[156]Department of Biostatistics and Epidemiology, Perelman School of Medicine, University of Pennsylvania, Philadelphia, PA, USA
[157]Department of Medicine, Perelman School of Medicine at the University of Pennsylvania, Philadelphia, PA, USA
[158]Center for Non-Communicable Diseases, Karachi, Pakistan
[159]National Institute for Health and Welfare, Helsinki, Finland
[160]Wellcome Sanger Institute, Wellcome Genome Campus, Hinxton, Cambridge, UK
[161]Deutsches Herzzentrum, München, Germany
[162]Technische Universität München, Germany
[163]Institute of Genetic Epidemiology, Department of Genetics and Pharmacology, Medical University of Innsbruck, 6020 Innsbruck, Austria
[164]Duke Molecular Physiology Institute, Durham, NC

[165]Division of Cardiovascular Medicine, Nashville VA Medical Center, Vanderbilt University School of Medicine, Nashville, TN, USA
[166]Division of Endocrinology, National University Hospital, Singapore
[167]NUS Saw Swee Hock School of Public Health, Singapore
[168]Channing Division of Network Medicine, Brigham and Women's Hospital, Boston, MA, USA
[169]Harvard Medical School, Boston, MA, USA
[170]Department of Psychiatry, Icahn School of Medicine at Mount Sinai, New York, NY, USA
[171]Department of Genetics and Genomic Sciences, Icahn School of Medicine at Mount Sinai, New York, NY, USA
[172]Institute for Genomics and Multiscale Biology, Icahn School of Medicine at Mount Sinai, New York, NY, USA
[173]The Wallenberg Laboratory/Department of Molecular and Clinical Medicine, Institute of Medicine, Gothenburg University and the Department of Cardiology, Sahlgrenska University Hospital, Gothenburg, Sweden
[174]Department of Cardiology, Wallenberg Center for Molecular Medicine and Lund University Diabetes Center, Clinical Sciences, Lund University and Skåne University Hospital, Lund, Sweden
[175]Institute of Clinical Medicine Neurology, University of Eastern Finad, Kuopio, Finland
[176]Sorbonne Université, INSERM, Centre de Recherche Saint-Antoine, CRSA, AP-HP, Saint Antoine Hospital, Gastroenterology department, F-75012 Paris, France
[177]INRA, UMR1319 Micalis & AgroParisTech, Jouy en Josas, France
[178]Paris Center for Microbiome Medicine, (PaCeMM) FHU, Paris, France
[179]Department of Twin Research and Genetic Epidemiology King's College London, London, UK
[180]The McDonnell Genome Institute at Washington University, Seattle, WA, USA
[181]Departments of Genetics and Psychiatry, University of North Carolina, Chapel Hill, NC, USA
[182]Saw Swee Hock School of Public Health National University of Singapore, National University Health System, Singapore

[183]Department of Medicine, Yong Loo Lin School of Medicine National University of Singapore, Singapore
[184]Duke-NUS Graduate Medical School, Singapore
[185]Life Sciences Institute, National University of Singapore, Singapore
[186]Department of Statistics and Applied Probability, National University of Singapore, Singapore
[187]Center for Behavioral Genomics, Department of Psychiatry, University of California, San Diego, CA, USA
[188]Institute of Genomic Medicine, University of California San Diego, San Diego, CA, USA
[189]Endocrinology, Abdominal Center, Helsinki University Hospital, Helsinki, Finland
[190]Institute of Genetics, Folkhalsan Research Center, Helsinki, Finland
[191]Juliet Keidan Institute of Pediatric Gastroenterology Shaare Zedek Medical Center, The Hebrew University of Jerusalem, Jerusalem, Israel
[192]Instituto de Investigaciones Biomédicas, UNAM, Mexico City, Mexico
[193]Instituto Nacional de Ciencias Médicas y Nutrición Salvador Zubirán, Mexico City, Mexico
[194]Department of Public Health Faculty of Medicine University of Helsinki, Helsinki, Finland
[195]Department of Psychiatry and Human Behavior, University of California Irvine, Irvine, CA, USA
[196]National Heart & Lung Institute & MRC London Institute of Medical Sciences, Imperial College, London, UK
[197]Cardiovascular Research Centre Royal Brompton & Harefield Hospitals, London, UK
[198]Radcliffe Department of Medicine, University of Oxford, Oxford, UK
[199]Department of Gastroenterology and Hepatology, University of Groningen and University Medical Center Groningen, Groningen, Netherlands
[200]Folkhälsan Institute of Genetics, Folkhälsan Research Center, Helsinki, Finland
[201]National Heart & Lung Institute and MRC London Institute of Medical Sciences, Imperial College London, London, UK

[202]Cardiovascular Research Centre, Royal Brompton & Harefield Hospitals NHS Trust, London, UK
[203]Department of Physiology and Biophysics, University of Mississippi Medical Center, Jackson, MS, USA
[204]Program in Infectious Disease and Microbiome, Broad Institute of MIT and Harvard, Cambridge, MA, USA
[205]Center for Computational and Integrative Biology, Massachusetts General Hospital, Boston, MA, USA

## GENOME AGGREGATION DATABASE CONSORTIUM FUNDING STATEMENTS

Matthew J. Bown: British Heart Foundation awards CS/14/2/30841 and RG/18/10/33842

Josée Dupuis: National Heart Lung and Blood Institute's Framingham Heart Study Contract (HHSNI); National Institute for Diabetes and Digestive and Kidney Diseases (NIDDK) R DK

Martti Färkkilä: State funding for university level health research

Laura D. Gauthier: Intel, Illumina

Stephen J. Glatt: U.S. NIMH Grant R MH

Leif Groop: The Academy of Finland and University of Helsinki: Center of Excellence for Complex Disease Genetics (grant number 312063 and 336822), Sigrid Jusélius Foundation; IMI 2 (grant No 115974 and 15881)

Mikko Hiltunen: Academy of Finland (grant 338182) Sigrid Jusélius Foundation the Strategic Neuroscience Funding of the University of Eastern Finland

Chaim Jalas: Bonei Olam

Jaakko Kaprio: Academy of Finland (grants 312073 and 336823)

Jacob McCauley: National Institute of Diabetes and Digestive and Kidney Disease Grant R01DK104844

Yukinori Okada: JSPS KAKENHI (19H01021, 20K21834), AMED (JP21km0405211, JP21ek0109413, JP21gm4010006, JP21km0405217, JP21ek0410075), JST Moonshot R&D (JPMJMS2021)

Michael J. Owen: Medical Research Council UK: Centre Grant No. MR/L010305/1, Program Grant No. G0800509

Aarno Palotie: the Academy of Finland Center of Excellence for Complex Disease Genetics (grant numbers 312074 and 336824) and Sigrid Jusélius Foundation

John D. Rioux: National Institute of Diabetes and Digestive and Kidney Diseases (NIDDK; DK062432), from the Canadian Institutes of Health (CIHR GPG 102170), from Genome Canada/Génome Québec (GPH-129341), and a Canada Research Chair (#230625)

Samuli Ripatti: the Academy of Finland Center of Excellence for Complex Disease Genetics (grant number) Sigrid Jusélius Foundation
Jerome I. Rotter: Trans-Omics in Precision Medicine (TOPMed) program was supported by the National Heart, Lung and Blood Institute (NHLBI). WGS for “NHLBI TOPMed: Multi-Ethnic Study of Atherosclerosis (MESA)” (phs001416.v1.p1) was performed at the Broad Institute of MIT and Harvard (3U54HG003067-13S1). Core support including centralized genomic read mapping and genotype calling, along with variant quality metrics and filtering were provided by the TOPMed Informatics Research Center (3R01HL-117626-02S1; contract HHSN268201800002I). Core support including phenotype harmonization, data management, sample-identity QC, and general program coordination were provided by the TOPMed Data Coordinating Center (R01HL-120393; U01HL-120393; contract HHSN268201800001I). We gratefully acknowledge the studies and participants who provided biological samples and data for MESA and TOPMed. JSK was supported by the Pulmonary Fibrosis Foundation Scholars Award and grant K23-HL-150301 from the NHLBI. MRA was supported by grant K23-HL-150280, AJP was supported by grant K23-HL-140199, and AM was supported by R01-HL131565 from the NHLBI. EJB was supported by grant K23-AR-075112 from the National Institute of Arthritis and Musculoskeletal and Skin Diseases.The MESA project is conducted and supported by the National Heart, Lung, and Blood Institute (NHLBI) in collaboration with MESA investigators. Support for MESA is provided by contracts 75N92020D00001, HHSN268201500003I, N01-HC-95159, 75N92020D00005, N01-HC-95160, 75N92020D00002, N01-HC-95161, 75N92020D00003, N01-HC-95162, 75N92020D00006, N01-HC-95163, 75N92020D00004, N01-HC-95164, 75N92020D00007, N01-HC-95165, N01-HC-95166, N01-HC-95167, N01-HC-95168, N01-HC-95169, UL1-TR-000040, UL1-TR-001079, and UL1-TR-001420. Also supported in part by the National Center for Advancing Translational Sciences, CTSI grant UL1TR001881, and the National Institute of Diabetes and Digestive and Kidney Disease Diabetes Research Center (DRC) grant DK063491 to the Southern California Diabetes Endocrinology Research Center
Edwin K. Silverman: NIH Grants U01 HL089856 and U01 HL089897

J. Gustav Smith: The Swedish Heart-Lung Foundation (2019-0526), the Swedish Research Council (2017-02554), the European Research Council (ERC-STG-2015-679242), Skåne University Hospital, governmental funding of clinical research within the Swedish National Health Service, a generous donation from the Knut and Alice Wallenberg foundation to the Wallenberg Center for Molecular Medicine in Lund, and funding from the Swedish Research Council (Linnaeus grant Dnr 349-2006-237, Strategic Research Area Exodiab Dnr 2009-1039) and Swedish Foundation for Strategic Research (Dnr IRC15-0067) to the Lund University Diabetes Center

Kent D. Taylor: Trans-Omics in Precision Medicine (TOPMed) program was supported by the National Heart, Lung and Blood Institute (NHLBI). WGS for “NHLBI TOPMed: Multi-Ethnic Study of Atherosclerosis (MESA)” (phs001416.v1.p1) was performed at the Broad Institute of MIT and Harvard (3U54HG003067-13S1). Core support including centralized genomic read mapping and genotype calling, along with variant quality metrics and filtering were provided by the TOPMed Informatics Research Center (3R01HL-117626-02S1; contract HHSN268201800002I). Core support including phenotype harmonization, data management, sample-identity QC, and general program coordination were provided by the TOPMed Data Coordinating Center (R01HL-120393; U01HL-120393; contract HHSN268201800001I). We gratefully acknowledge the studies and participants who provided biological samples and data for MESA and TOPMed.

JSK was supported by the Pulmonary Fibrosis Foundation Scholars Award and grant K23-HL-150301 from the NHLBI. MRA was supported by grant K23-HL-150280, AJP was supported by grant K23-HL-140199, and AM was supported by R01-HL131565 from the NHLBI. EJB was supported by grant K23-AR-075112 from the National Institute of Arthritis and Musculoskeletal and Skin Diseases.The MESA project is conducted and supported by the National Heart, Lung, and Blood Institute (NHLBI) in collaboration with MESA investigators. Support for MESA is provided by contracts 75N92020D00001, HHSN268201500003I, N01-HC-95159, 75N92020D00005, N01-HC-95160, 75N92020D00002, N01-HC-95161, 75N92020D00003, N01-HC-95162, 75N92020D00006, N01-HC-95163, 75N92020D00004, N01-HC-95164, 75N92020D00007, N01-HC-95165, N01-HC-95166, N01-HC-95167, N01-HC-95168, N01-HC-95169, UL1-TR-000040, UL1-TR-001079, and UL1-TR-001420. Also

supported in part by the National Center for Advancing Translational Sciences, CTSI grant UL1TR001881, and the National Institute of Diabetes and Digestive and Kidney Disease Diabetes Research Center (DRC) grant DK063491 to the Southern California Diabetes Endocrinology Research Center

Tiinamaija Tuomi: The Academy of Finland and University of Helsinki: Center of Excellence for Complex Disease Genetics (grant number 312072 and 336826), Folkhalsan Research Foundation, Helsinki University Hospital, Ollqvist Foundation, Liv och Halsa foundation; NovoNordisk Foundation

Teresa Tusie-Luna: CONACyT Project 312688

James S. Ware: Wellcome Trust [107469/Z/15/Z], Medical Research Council (UK), NIHR Imperial College Biomedical Research Centre

Rinse K. Weersma: The Lifelines Biobank initiative has been made possible by subsidy from the Dutch Ministry of Health Welfare and Sport the Dutch Ministry of Economic Affairs the University Medical Centre Groningen (UMCG the Netherlands) the University of Groningen and the Northern Provinces of the Netherlands